\documentclass[final,3p,times,review]{elsarticle}

\biboptions{numbers,sort&compress}

\usepackage{framed,multirow}
\usepackage[fleqn]{amsmath}
\usepackage{amssymb}
\usepackage{latexsym}
\usepackage{url}
\usepackage{xcolor}
\usepackage{graphicx}
\usepackage{float}
\usepackage{hyperref}
\usepackage{booktabs}
\usepackage{tabularx}
\usepackage{afterpage}
\usepackage{lipsum}
\usepackage{pdfpages}
\usepackage{CJKutf8}
\usepackage{caption}
\usepackage{wrapfig}
\usepackage{tikz}
\usepackage{array}
\usepackage{amsbsy}
\usepackage{bm}
\usepackage{fancyhdr}
\journal{Expert Systems With Applications}

\begin{document}

\begin{frontmatter}



\title{DEFN: Dual-Encoder Fourier Group Harmonics Network for Three-Dimensional Indistinct-Boundary Object Segmentation}

%

\author[1]{Xiaohua Jiang \fnref{fn1}}
\author[4]{Yihao Guo \fnref{fn1}}
\author[4]{Jian Huang \fnref{fn1}}
\author[1]{Yuting Wu}
\author[1]{Meiyi Luo}
\author[2]{Zhaoyang Xu}
\author[3]{Qianni Zhang}
\author[4]{Xingru Huang}
\author[4]{Hong He}
\author[4]{Shaowei Jiang\corref{cor1}}
\author[1]{Jing Ye\corref{cor1}}
\author[1]{Mang Xiao\corref{cor1}} \ead{joelxm@zju.edu.cn}

\fntext[fn1]{These authors contributed equally to this work.}
\cortext[cor1]{Corresponding author.}
      
\affiliation[1]{organization={The department of Otolaryngology Head and Neck surgery, Sir Run Run Shaw Hospital, Affiliated to Zhejiang University School of Medicine.},
                city={Hangzhou},
                country={China}}
\affiliation[2]{organization={Department of Paediatrics, University of Cambridge},
                postcode={CB2 1TN}, 
                city={Cambridge},
                country={UK}}
\affiliation[3]{organization={School of Electronic Engineering and Computer Science, Queen Mary University of London},
                postcode={E3 4BL}, 
                city={London},
                country={UK}}
\affiliation[4]{organization={Hangzhou Dianzi University},
                city={Hangzhou},
                country={China}}
                

\begin{abstract}
The precise spatial and quantitative delineation of indistinct-boundary medical objects is paramount for the accuracy of diagnostic protocols, efficacy of surgical interventions, and reliability of postoperative assessments. Despite their significance, the effective segmentation and instantaneous three-dimensional reconstruction are significantly impeded by the paucity of representative samples in available datasets and noise artifacts. To surmount these challenges, we introduced Stochastic Defect Injection (SDi) to augment the representational diversity of challenging indistinct-boundary objects within training corpora. Consequently, we propose the Dual-Encoder Fourier Group Harmonics Network (DEFN) to tailor noise filtration, amplify detailed feature recognition, and bolster representation across diverse medical imaging scenarios. By incorporating Dynamic Weight Composing (DWC) loss dynamically adjusts model’s focus based on training progression, DEFN achieves SOTA performance on the OIMHS public dataset, showcasing effectiveness in indistinct boundary contexts. Source code for DEFN is available at: \href{https://github.com/IMOP-lab/DEFN-pytorch}{https://github.com/IMOP-lab/DEFN-pytorch}.
\end{abstract}







\begin{keyword}
Three-demensional segmentation\sep Indistinct-boundary\sep Hypopharyngeal cancer\sep Augmentative stochasticity\sep Fourier harmonic\sep Spatio-temporal attention
\end{keyword}

\end{frontmatter}


\section{Introduction}
\begin{figure*}[!t]
    \centering
    \includegraphics[width= 0.9\linewidth]{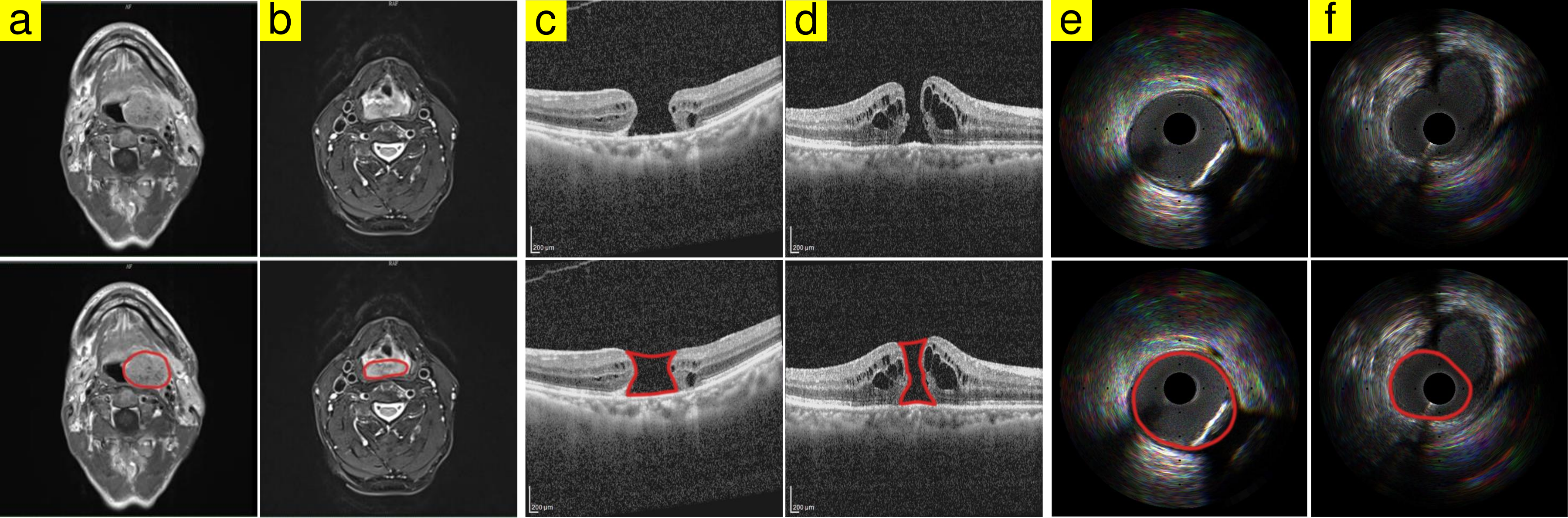}
    \caption{Indistinct-boundary challenges in medical imaging segmentation. (a, b) Hypopharyngeal cancer presents difficulties in boundary delineation due to submucosal growth. (c, d) Macular holes pose segmentation challenges with their inferred upper boundaries. (e) Calcific shadows in coronary artery IVUS imaging obscure arterial boundaries. (f) Side-branches further complicate segmentation. }
    \label{fig01}
\end{figure*}

In the domain of diagnostic medical imaging, the delineation of indistinct-boundary entities remains a formidable challenge, particularly salient in pathologies where the nascent detection is pivotal for effective therapeutic interventions. This elusiveness significantly impedes automated segmentation processes, essential for quantifying the extent and morphology of the pathology, thus complicating early diagnosis and intervention strategies. Among the most compelling clinical illustrations is the challenge posed by hypopharyngeal cancer, where early detection is arduous and seldom achieved. The disease is predominantly diagnosed in its advanced stages due to its propensity for submucosal growth, which blends seamlessly into the surrounding pharyngeal tissues in MR images, thereby obscuring clear boundary demarcation (Fig.\ref{fig01} a,b). This characteristic complicates the automated segmentation and quantification necessary for early therapeutic intervention, adversely impacting overall survival rates. Similarly, the segmentation of macular holes introduces additional complexities. The upper boundary of a macular hole is typically inferred rather than directly observed, as the hole itself is essentially a void with no physical upper edge (Fig.\ref{fig01} c,d). This necessitates predictive modeling to define inherently theoretical boundaries, further complicating the segmentation process and underscoring the need for advanced computational approaches. In the case of coronary artery imaging using Intravascular Ultrasound (IVUS), the challenges proliferate with the presence of calcific shadows (Fig.\ref{fig01} e) and side-branch (Fig.\ref{fig01} f), while calcifications block ultrasound signals, creating shadowed regions behind them that obscure the arterial boundaries. The task then extends beyond mere imaging to involve predictive localization of the arterial walls within these optically occluded zones and side-branches to create a conceptual border, which requires profound analytical ingenuity to execute accurately.

The exigency of devising sophisticated methods for the stable segmentation of indistinct-boundary medical objects is underscored by their pivotal role in three-dimensional medical reconstructions and precise lesion quantification. Contemporary segmentation techniques frequently falter due to the dual constraints of sparse sample availability and pronounced noise artifacts within imaging data. These constraints drastically curtail the efficacy of existing models, precipitating a crucial need for innovative solutions capable of transcending these limitations.

Due to the constraints imposed by limited data availability, we specifically selected the task of macular hole segmentation to evaluate the performance of our proposed methods. Macular degeneration represents a formidable adversary and stands as a leading cause of vision impairment alongside glaucoma and cataracts \citep{coleman2010impact}. This condition is characterized by the deterioration of the macula, a region densely populated with photoreceptor cells in the retina, and necessitates early intervention to forestall irreversible central vision loss \citep{mitchell2018age}. At the heart of this degenerative condition is the phenomenon of the macular hole, whose prompt and precise assessment is vital for determining an efficacious treatment strategy \citep{ho1998macular}. The task of accurately discerning the spatial coordinates of the macular hole is of paramount importance as it informs the classification based on size and shape, ultimately guiding the selection of a suitable therapeutic approach \citep{ullrich2002macular}. Surgical interventions targeting the macular hole require a nuanced determination of the filler type and dosage, underscoring the necessity for a preoperative understanding of the hole's dimensions and location to tailor a surgical plan that optimizes outcomes and procedural efficiency\citep{salter2012macular}. As the treatment unfolds, metrics pertaining to the macular hole serve as a crucial benchmark for evaluating the efficacy of the intervention.

\begin{figure*}[!t]
    \centering
    \includegraphics[width=\linewidth]{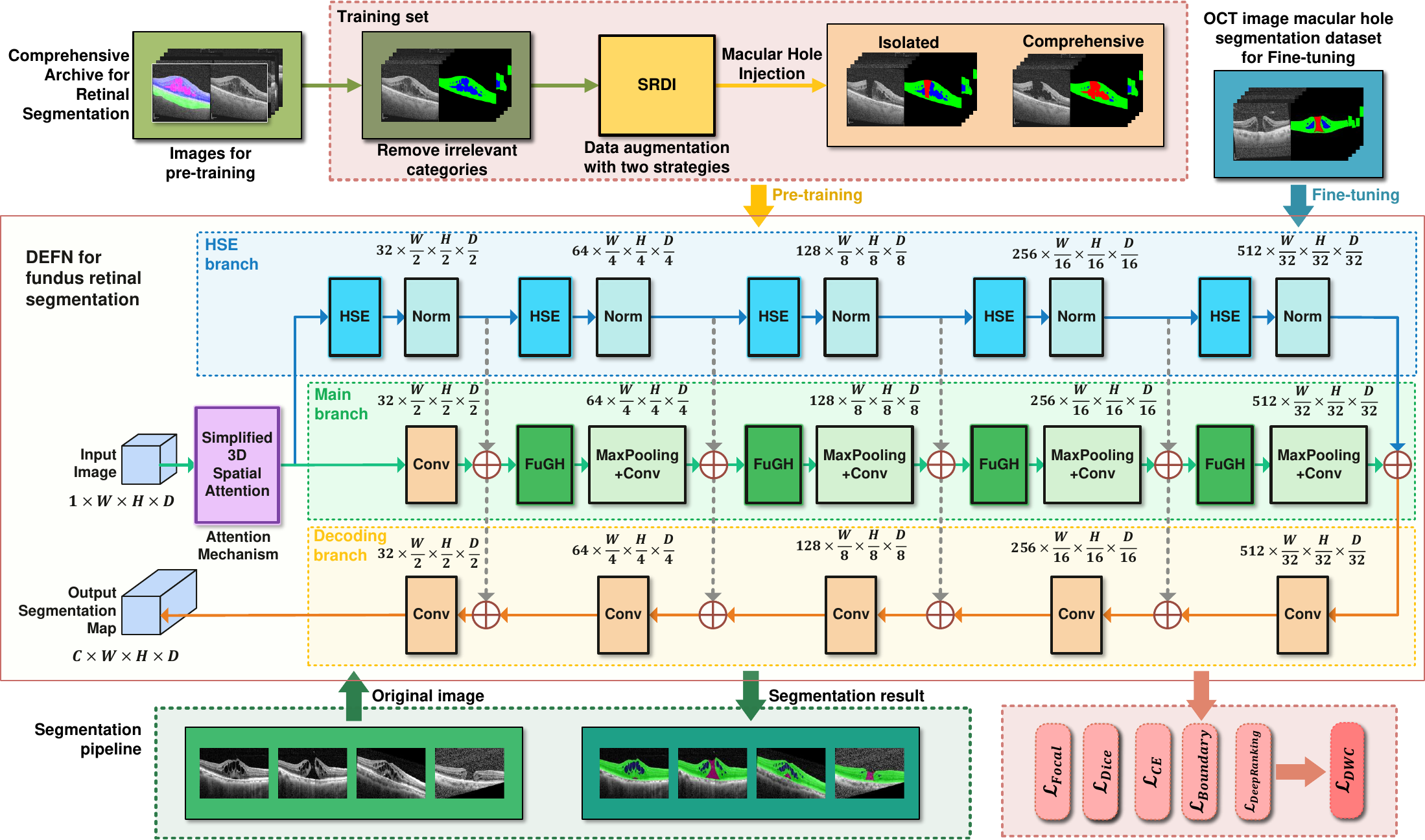}
    \caption{Schematic representation of the project workflow encompassing a data augmentation pipeline utilizing SDi, a data amplification technology, the pre-training and fine-tuning phases for the DEFN, along with the segmentation pipeline. The SDi includes two strategies: isolated and comprehensive injection. The DWC, a dynamic weight composing network optimization strategy, is integrated into this workflow to enhance the model performance.}
    \label{fig02}
\end{figure*}

\afterpage{%
  \footnotetext[2]{The source code is publicly available at: \href{https://github.com/IMOP-lab/DEFN-pytorch}{https://github.com/IMOP-lab/DEFN-pytorch}.}
}

However, the current landscape of accurately segmenting ocular structures, including the macular hole, especially in environments with high noise interference and conceptual boundaries, is marked by inadequacies. A notable challenge is the scarcity of macular hole samples, which hampers the efficacy of data-driven analysis methods due to insufficient training data. Moreover, the precise segmentation of the macular hole is complicated by disturbances such as interference with the cellular proliferation at the edges of the retina in the macular region and the presence of intra-vitreous impurities. The commonality of vitreous turbidity as a form of noise further detracts from the quality of imaging, impeding segmentation performance \citep{steel2013idiopathic}. Additionally, the upper edge of the macular hole often lacks distinct features, complicating the prediction of its upper boundary. The differentiation between the macular hole and subretinal fluid in image sequences poses another layer of complexity, exacerbating the challenge of continuity determination.

To surmount these challenges, we proposed the Dual-Encoder Fourier Group Harmonics Network (DEFN\textsuperscript{2}), a novel segmentation network featuring the Fourier Group Harmonics (FuGH) module, the Simplified 3D Spatial Attention (S3DSA) module, and the Harmonic Squeeze-and-Excitation Module (HSE). This network is intricately designed to enhance the segmentation accuracy and robustness of ocular anatomical structures, including the macular hole among other pivotal elements within the visual system. Through the employment of the FuGH module, we enhance the recognition of periodic patterns via frequency-domain group convolution, mitigating noise influence and bolstering segmentation accuracy. The S3DSA module enhances the model's focus on critical areas by adjusting the weight of spatial region information, thereby markedly improving the identification accuracy of challenging areas such as the macular hole and edema. The incorporation of the HSE module facilitates the precise emphasis on critical features through the learning of complex inter-channel weight patterns.

To architect the DEFN for retinal segmentation, our research embarked on an extensive five-year endeavor to compile the Comprehensive Archive for Retinal Segmentation (CARS-30k), the most expansive dataset of its kind to date, meticulously annotated by seasoned clinicians to delineate macular edema, along with retinal and optic disc structures. This dataset formed the basis for implementing a data augmentation technique, Stochastic Defect Injection (SDi), to artificially increase the representation of macular holes in CARS-30k. The evaluation of DEFN with a novel network optimization strategy, Dynamic Weight Composing (DWC), against 13 baseline models on the publicly available OIMHS \cite{ye2023oimhs} dataset underscores its superiority in segmenting macular holes and other key ocular components, further enhanced by a three-dimensional structure reconstruction methodology for detailed visualization of the fundus structure. The overall structure is illustrated in Fig.\ref{fig02}.


Our study's primary contributions are summarized as follows:

1. We introduced the SDi data augmentation technique to address the limitations associated with the scarcity of indistinct-boundary object data and sequence sizes. Additionally, this method applies to the augmentation of other small datasets.

2. We proposed a novel segmentation network DEFN, tailored for the segmentation of medical images characterized by substantial noise. DEFN comprises three pivotal modules: FuGH, S3DSA, and HSE, construct an enhanced frequency domain representation that facilitates a broader global receptive field and enables noise reduction.

3. We introduced the concept of DWC, a novel network optimization strategy that allows for the dynamic fusion and adjustment of loss functions tailored to the varying requirements of different training stages. This strategy effectively elevates the segmentation accuracy and robustness of indistinct-boundary objects.

4. We developed a comprehensive 3D reconstruction scheme based on the segmentation results, enabling precise modeling and quantitative analysis of retinal structures.

More background and related works are exhaustively discussed in Chapter 2. All methods mentioned above are described in detail in Chapter 3. Experimental data is provided in Chapter 4, along with comprehensive evaluation results. The conclusion is given and concluded in chapter 5.

\section{Related work}
\subsection{Medical image segmentation}
In the field of medical image segmentation, particularly concerning the delineation of structures within retinal imagery, there has been a marked evolution from traditional image processing techniques towards deep learning methodologies, especially Convolutional Neural Networks. This transition has catalyzed significant advancements in accuracy and efficiency, enabling more nuanced analysis and interpretation of medical images. Among the array of architectures that have emerged, the 3D U-Net \citep{cciccek20163d} stands out for its efficacy in volumetric segmentation, demonstrating remarkable versatility across various applications. The advent of deep learning, epitomized by neural networks such as V-Net \citep{milletari2016v}, Ori-Net \citep{JIANG2024121905}, and CS3DEA-Net\citep{RAJESH2024122509}, has redefined benchmarks in medical image segmentation tasks, showcasing enhanced performance and robustness.

Innovations within this domain have not been limited to architectural enhancements alone. Focused efforts have been directed towards addressing computational constraints and improving segmentation precision, particularly for challenging scenarios like imbalanced classes or the delineation of intricate structures such as vessel-like formations in 3D medical images. Notable among these innovations are compact convolutional networks designed for real-time applications on clinically constrained computational platforms \citep{dai2022can3d} and edge-reinforced neural networks aimed at refining the segmentation of crisp edges within vessel-like structures \citep{xia20223d}. Moreover, the evolution of architectural frameworks such as MPFC-Net \citep{WU2024123430} introduces a pioneering network design intended to counteract semantic loss in continuous pooling and convolution by employing multi-perspective feature compensation and serialized multi-scale blocks, thereby demonstrating robust performance.

The segmentation of retinal structures, a critical component of ophthalmological imaging, has similarly benefited from these advancements. Architectures utilizing multi-encoder-decoder configurations in deep convolutional neural networks have significantly improved the segmentation of retinal vasculature \citep{CHALA2021115459}. Innovations such as stimulus-guided adaptive transformer networks \citep{lin2023stimulus} and boundary-aware context neural networks \citep{wang2022boundary} represent significant strides toward achieving precise segmentation in ophthalmological images. These developments not only underscore the rapid evolution of medical image segmentation techniques but also highlight the potential of deep learning in enhancing diagnostic accuracy and facilitating better patient outcomes.

Due to insufficient data, data augmentation techniques to tackle the inherent complexities of retinal imaging in optical coherence tomography (OCT) have only been preliminarily explored. Mamta Juneja et al. \citep{JUNEJA2022117202} proposed a computer-assisted diagnostic system that integrates machine learning and deep learning, utilizing fundus imaging and OCT for the classification and monitoring of glaucoma. The work of İsmail Kayadibi et al.  \citep{KAYADIBI2023120617} developed a hybrid Retinal Fine-Tuned Convolutional Neural Network (R-FTCNN) combined with Principal Component Analysis (PCA), significantly enhancing the accuracy of detecting retinal diseases from OCT images through optimized feature extraction and classification processes. Building upon these advancements, Wang et al. \citep{WANG2024123496} introduced an Anatomy and multi-label semantic consistency network (AMSC-Net), which employs a minimal amount of labeled data and a heterogeneous architecture consistency strategy to improve segmentation performance and robustness, demonstrating exceptional results and adaptability across datasets. These developments in data enhancement methods, especially within the context of OCT segmentation, not only enhance the accuracy and efficiency of segmentation tasks but also make substantial contributions to the broader domain of medical image analysis, facilitating more precise and dependable diagnostic interpretations.

\subsection{Frequency domain information fusion}
The assimilation of frequency domain transformations, specifically the fast Fourier transform (FFT) and wavelet transform, within neural networks represents a notable development in enhancing computational efficiency and accuracy for a range of tasks. The incorporation of FFT has been extensively examined for its potential to augment the performance of CNNs, yielding effective improvements in both accuracy and computational expediency.

Recent progress preliminary tested the applicability of FFT within deep learning frameworks. The seminal contributions by 
Yuan et al. \citep{YUAN2024110428} notably enhanced the segmentation accuracy in low-contrast images by incorporating FFT into their LCSeg-Net model, which effectively suppresses noise and emphasizes crucial features within the frequency domain, substantially improving the robustness and clarity of segmentation results. Wang et al. \citep{WANG2022108794} significantly improved image segmentation efficiency by incorporating the FFT in their Iterative Convolution–Thresholding Method (ICTM), enabling faster and simpler processing compared to traditional methods. Furthermore, the novel methodology introduced by Guibas et al. \cite{guibas2021adaptive}, which integrates FFT with transformer models, showcases the technique's capacity to improve token mixing efficiency, underscoring the transformative potential of frequency domain information fusion in deep learning landscapes.

\subsection{Loss function for medical image segmentation}
The strategic selection of appropriate loss functions is crucial for the optimization of deep learning architectures. Within the extensive array of loss functions, the dice loss ($\mathcal{L}_{Dice}$) \citep{milletari2016v} and the cross-entropy loss ($\mathcal{L}_{CE}$) are universally recognized for their straightforwardness and efficacy across a diverse range of applications. In contrast, the focal loss ($\mathcal{L}_{Focal}$) \citep{lin2017focal} is designed to address the prevalent challenge of class imbalance by diminishing the influence of correctly classified examples, thereby directing the model's learning focus towards more complex instances. The introduction of unified focal loss ($\mathcal{L}_{Unified focal}$) \citep{yeung2022unified} provides an integrative framework that synthesizes the strengths of both $\mathcal{L}_{Dice}$ and $\mathcal{L}_{CE}$, offering a comprehensive and resilient methodology for loss function development. Furthermore, the boundary loss ($\mathcal{L}_{Boundary}$) \citep{kervadec2019boundary} emphasizes the importance of accurate boundary identification in segmentation endeavors, underscoring the bespoke requirements incumbent upon medical imaging tasks. 

Although these disparate loss functions excel in particular aspects, they frequently fall short in furnishing a universally superior solution across the spectrum of segmentation challenges. This elucidates the imperative for meticulous selection and, where necessary, the bespoke amalgamation of loss functions to augment model efficacy within the intricate realm of medical image segmentation.

\section{Method}
The accurate delineation of boundary-clear objects is largely attributed to its relatively simple structure and pronounced features. However, the segmentation in blurred regions or conceptual boundaries, such as macular hole and macular edema, which often present with indistinct boundaries and are prone to confusion with vitreous opacities, remains a formidable challenge. The scarcity of datasets specific to the macular hole, especially with the limited sample size and sequences containing single samples, further compounds this challenge and hinders the advancement of model performance in segmenting the macular hole. Moreover, conventional loss functions employed in medical image segmentation tend to exhibit performance improvements confined to specific regions, such as boundary areas, while preserving limitations in the overall segmentation performance. To address these issues, we introduce an innovative data augmentation technique, Stochastic Defect Injection (SDi), and a segmentation network, Dual-Encoder Fourier Group Harmonics Network (DEFN). The SDi technique mitigates the issue of macular hole sample scarcity through random injection of simulated macular holes and sequence expansion, thereby enriching the training dataset. The DEFN network integrates three key modules: Fourier Group Harmonics (FuGH), Simplified 3D Spatial Attention (S3DSA), and Harmonic Squeeze-and-Excitation Module (HSE), enhancing the network's capability to address noise, blurred labels, and complex structural hierarchies in OCT images, consequently elevating the segmentation accuracy and robustness for the macular hole and other key structures. Additionally, a network optimization strategy, Dynamic Weight Composing (DWC), is introduced to fuse multiple loss functions, further refining the precision in segmenting challenging regions such as the macular hole and macular edema. This architecture, as detailed in subsequent sections, is schematically illustrated in Fig.\ref{fig01}.

\subsection{Stochastic Defect Injection}
The endeavor to identify and reconstruct conceptual Boundaries such as macular hole conditions presents several challenges. A primary challenge involves the sparsity of datasets, where each sample comprises a limited number of slices, resulting in a limited number of instances of macular holes. The scarcity of instances commonly exists, even within the most extensive publicly accessible datasets, significantly impedes the performance of trained deep learning networks for macular hole segmentation.

To facilitate effective model training and validation, a data augmentation technique termed SDi is proposed, which consists of two steps, sequence expansion and simulative injection. Through sequence expansion, specifically for 3D fundus sequences that initially do not contain macular holes, each sample undergoes augmentation via bilinear interpolation to increase the total number of slices to a predefined count.

\begin{figure}[!h]
\centering
\includegraphics[width=0.9\linewidth]{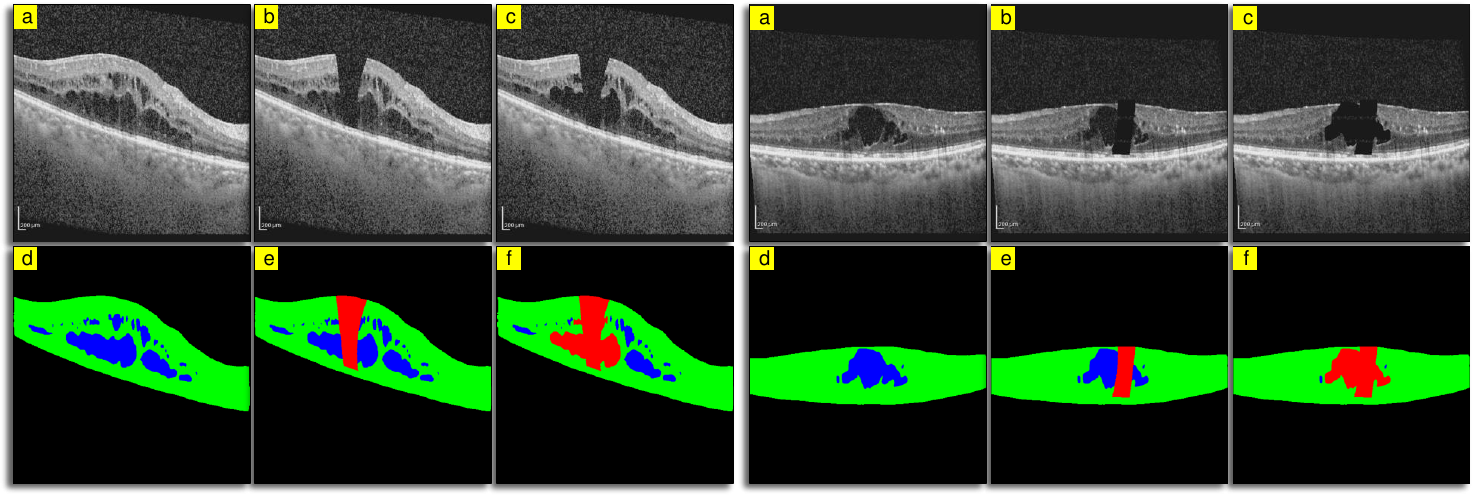}
\caption{Results of employing SDi: (a) original image, (b) image processed with the isolated injection strategy, (c) image processed with the comprehensive injection strategy, (d) corresponding mask for (a), (e) corresponding mask for (b), and (f) corresponding mask for (c). Different colors in masks represent different eye conditions: the green area represents the retina, the red area represents the macular hole, and the blue area represents the macular edema.}
\label{fig03}
\end{figure}

Subsequently, simulated defects are randomly injected into the expanded dataset, thus artificially boosting the presence of macular holes in the dataset. Nevertheless, indiscriminate injection of simulated macular holes may lead to occluding existing macular edema regions, potentially resulting in overlap between macular holes and macular edema. To mitigate this risk, SDi incorporates two injection strategies: isolated injection modifies all pixel classifications within the injection zone to the macular hole category while preserving the integrity of other regions, and comprehensive injection additionally reclassifies all pixels within edema regions connected to the injection area to the macular hole category. The effect of these strategies is graphically explicated in Fig.\ref{fig03}.

In the isolated injection strategy, the first step involves calculating a random distortion strength and the centroid coordinates of the voxel. For each slice within the voxel, if the slice is partitioned into halves along the y-axis, the primary and secondary radii along the y-axis at the current coordinates are computed. Utilizing the radii, distortion strength, and centroid coordinates, areas representative of macular holes are marked, as illustrated in parts b and e of Fig.\ref{fig03}.

The comprehensive injection strategy follows the same initial steps as the isolated strategy but additionally employs a depth-first search algorithm \citep{tarjan1972depth} to label all pixels within edema regions directly connected to the injected macular hole area as the macular hole category, as depicted in parts c and f of Fig.\ref{fig03}. 

Both the isolated and comprehensive strategies establish a 15-pixel isolation boundary post-simulation of macular hole regions. Regardless of the selected strategy for macular hole injection, the corresponding images can be reconstructed inversely from the segmentation masks through a similar process. This includes mask preprocessing to identify the macular hole areas in the corresponding images, followed by selectively cropping the image background. After stitching the cropped area and applying Gaussian blurring, it is overlaid onto the macular hole area of the original image. In this way, a fundus OCT sequence enriched with realistic macular holes is generated.

This approach to data augmentation via SDi plays a pivotal role in enhancing the quality of model training and validation, thereby addressing the critical challenge posed by the scarcity of training samples.

\begin{figure*}[!t]
    \centering
    \includegraphics[width=\linewidth]{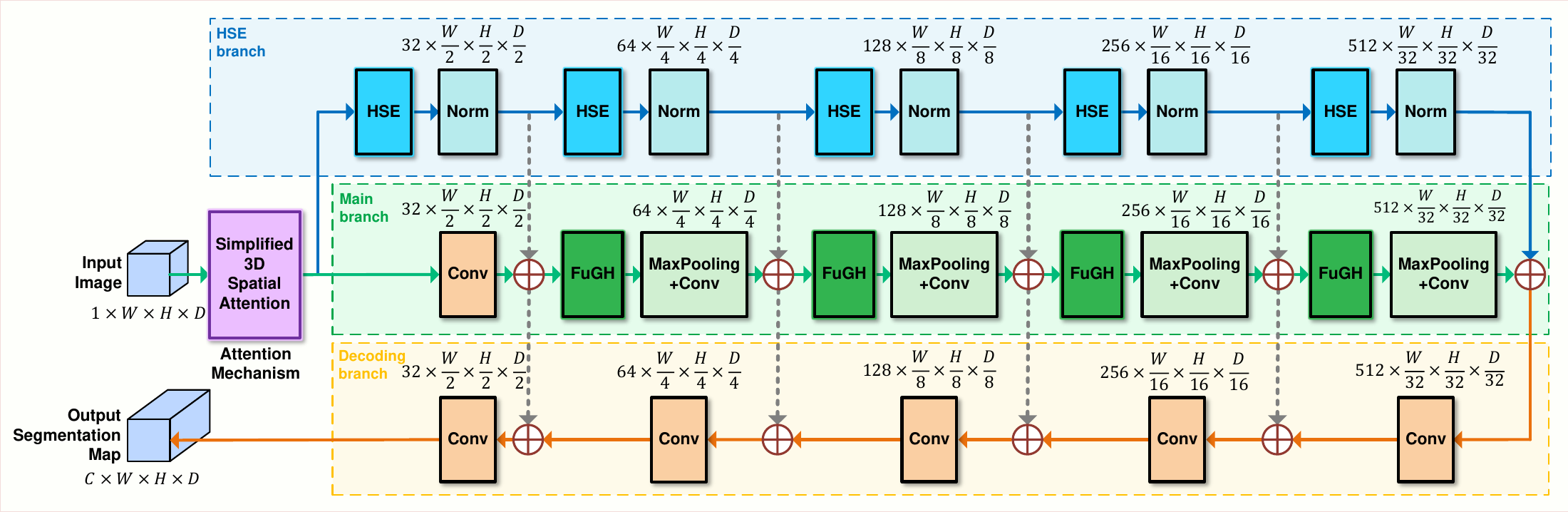}
    \caption{Schematic representation of the DEFN architecture, encompassing a main branch, an HSE branch, and a decoding branch. Within the HSE branch, five consecutive HSE modules are interlaced with Norm operations. The Main branch is structured with a Conv layer, followed by four consecutive FuGH modules interlaced with MaxPooling operations. The Decoding branch comprises five upsampling layers. Integral to this architecture is a Simplified 3D Spatial Attention mechanism, honing the model's focus on relevant spatial features.}
    \label{fig04}
\end{figure*}

\subsection{Dual-Encoder Fourier Group Harmonics Network}
Segmentation of fuzzy 3D objects not only requires precise localization and labeling of target regions within complex hierarchical structures, but also requires robustness against environmental noise and other interfering factors. To address these challenges, we propose the DEFN for 3D segmentation in noisy environments. As illustrated in Fig.\ref{fig04}, this network employs a main branch, an HSE branch, and a decoding branch, incorporating three key modules: FuGH, S3DSA, and HSE.

The FuGH module aims to recognize periodic structures and symmetric structures while reducing noise interference. Compared to conventional CNNs, FuGH improves the recognition of periodic patterns through frequency domain grouped convolutions, thereby reducing noise effects and improving segmentation accuracy. The S3DSA module is designed to address the issues of uneven anatomical layers and blurred labels in 3D OCT images. It adaptively adjusts the weights of spatial information distributions in the input data by increasing the weights of the macular hole and macular edema regions, promoting sensitivity in identifying these difficult-to-label regions. The HSE module enhances dependencies between channels by learning complex inter-channel weighting patterns, so as to highlight the important features and further enhance the discrimination ability of the model.

Overall, these three modules work together to effectively address noise, blurred labels, and hierarchical complexity in 3D-sequence segmentation. Experimental results validate that the integration of these modules enables the DEFN to excel at segmenting the intraretinal macular hole.

\subsubsection{Network structure}
To extract descriptive and distinctive features, the intricacy inherent in segmenting irregularly distributed macular edema and prognosticating the superior confines of the macular hole is magnified by noise perturbations attributable to factors such as vitreous opacity and environmental disturbances. These disturbances complicate the extraction of salient features, thereby negatively affecting the segmentation fidelity of macular edema boundaries and the prediction accuracy of the macular hole's upper demarcation. To reduce noise and augment feature extraction, the network's encoder is divided into two branches. The main branch employs the FuGH module alongside CNN downsampling, wherein the former processes frequency domain information to mitigate noise interference, whilst the latter is employed to extract profound spatial features. Concurrently, in the HSE branch, the core component is the HSE module. By integrating a FuGH module in its process, the HSE module achieves an expansive field of view within the frequency domain, enhancing the weights of positive feature channels while suppressing those considered negative. Subsequent to feature input into the primary channel and ensuing CNN downsampling, each iteration doubling feature channels and halving feature dimensions, the decoder synergizes the output from the preceding level with encoder-correspondent features through upsampling. In this way, the number of feature channels is reduced by half whilst the dimension is doubled. After four upsampling layers, a Conv layer transmutes the channels to align with the number of categories, with each channel focusing on the segmentation outcome per class. The comprehensive network structure is elucidated in Fig.\ref{fig04}.

\subsubsection{Fourier Group Harmonics}
\begin{figure}[!h] 
    \centering
    \includegraphics[width=0.6\linewidth]{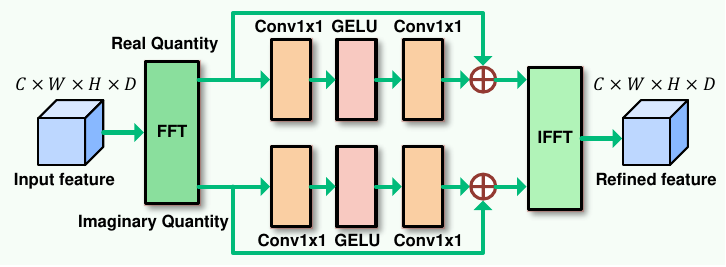}
    \caption{Schematic representation of the FuGH module, initiating with an FFT operation that segregates the input into real and imaginary components. These components are then sequentially channeled through a Conv layer, a GELU activation function, and a subsequent Conv layer on both upper and lower pathways. Following these processing stages, an IFFT operation is executed to yield the final output.}
    \label{fig05}
\end{figure}

Noise is also a major cause of creating ambiguous boundaries, interference in noise images poses additional challenges to segmentation. Conventional methods usually rely on spatial domain convolutions, and are therefore subject to noise, leading to inferior model robustness. To tackle this problem, the Fourier Group Harmonics (FuGH) module offers a feasible approach within CNN architectures. It employs the FFT to transit the model's operational domain from spatial to frequency. This paradigm shift enables a refined comprehension of data frequency characteristics, facilitating the identification and subsequent filtering of extraneous noise in both high-frequency and low-frequency domains. In addition, the FuGH module also incorporates group-based convolutional operations within the frequency domain, offering effective feature extraction and manipulation across specific frequency spectra. This process improves the model's capability to distinguish periodic patterns inherent in the data, thereby enhancing segmentation accuracy. Moreover, the group-based operations streamline the model by reducing the number of parameters and computational complexity. The structural diagram of the FuGH module is shown in Fig.\ref{fig05}.

For an input feature $x$, we first perform an FFT operation to obtain its representation in the frequency domain:
\begin{flalign}
&x_{fft} = FFT(x)&
\end{flalign}

Hereafter, we separately process the real and imaginary components of the data in the frequency domain. For both components, we apply two grouped convolutional operations with a GELU activation function inserted in between, and add a residual connection:
\begin{flalign}
&y_{fft_{real}}=Conv_2(GELU(Conv_1(x_{fft_{real}})))+x_{fft_{real}}& \\
&y_{fft_{imag}}=Conv_2(GELU(Conv_1(x_{fft_{imag}})))+x_{fft_{imag}}&    
\end{flalign}

Here, $x_{fft_{real}}$ and $x_{fft_{imag}}$ denote the real and imaginary components of the $x_{fft}$, respectively. $Conv_1$ and $Conv_2$ denote the first and second grouped convolutional operations, respectively.
In the last, the processed real and imaginary components are combined, and inverse fast Fourier transform (IFFT) is performed to resume the spatial domain representation $y_{out}$:
\begin{flalign}
&y_{out}=IFFT(Complex(y_{fft_{real}},y_{fft_{imag}}))&
\end{flalign}

The FuGH module provides the network with a new dimension to understand and extract key information from the data, especially when dealing with data involving clear and complex frequency patterns. 

\subsubsection{Simplified 3D Spatial Attention}
For indistinct-boundary objects , it is difficult to extract features due to their blurred boundaries against the retina and their appearance, which is highly similar to vitreous opacity. Segmenting the macular hole is a key focus as well as a challenge in fundus OCT segmentation, and effective feature extraction is the key to solving this task.

\begin{figure}[!h]
  \centering
  \includegraphics[width=0.6\linewidth]{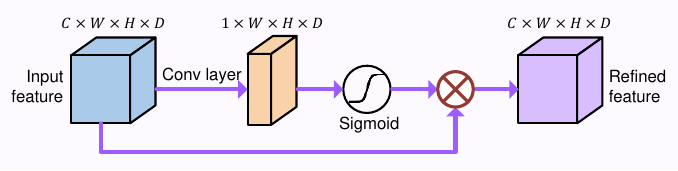}
  \caption{Schematic representation of the S3DSA module, which initially channels the input through a 3D Conv layer, followed by a Sigmoid activation function, and employs a skip connection. This configuration facilitates effective feature extraction while maintaining a simple structure of the framework.}
  \label{fig06}
\end{figure}

Attention mechanisms are widely applied in medical image segmentation tasks to enhance feature extraction. Through experimenting with multiple attention mechanisms, the best overall performance was achieved with spatial attention \citep{jaderberg2015spatial}, likely because spatial attention focuses on activating specific spatial regions, which is particularly effective for addressing problems such as vitreous opacities and background noise around the macular hole. Therefore, we develop based on spatial attention and further simplify it, yielding the S3DSA module as shown in Fig.\ref{fig06}.

The S3DSA module processes the original input features through a 3D Conv layer, generating attention weights. These weights are subsequently normalized within the range of [0,1] using a sigmoid activation function. The sigmoid activation function is formulated as follows:
\begin{flalign}
&S(x)=\frac{1}{1+exp(-x)}&
\end{flalign}

Finally, the original input features $x$ and the attention weights $Att_{s}$ normalized by the sigmoid function undergo an element-wise multiplication to obtain the weighted output. This can be represented by the formula:
\begin{flalign}
&{Output}=x\odot{Att}_{s}&
\end{flalign}

By preprocessing the input using S3DSA, the network is able to focus attentional resources on more important macular holes and macular edema regions within the input volume. This helps to improve segmentation quality. Furthermore, S3DSA simplifies the underlying spatial attention mechanism by removing unnecessary operations and optimizing the design for higher computational efficiency.

\subsubsection{Harmonic Squeeze-and-Excitation Module}
\begin{figure*}[!t]
    \centering
    \includegraphics[width=0.8\linewidth]{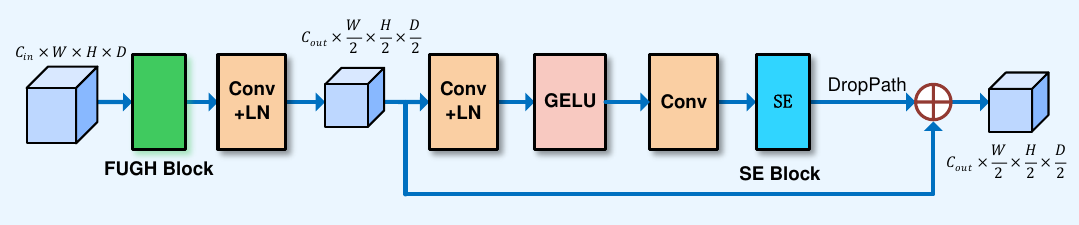}
    \caption{Schematic representation of the HSE module, which employs the FuGH block for initial spatial and frequency domain information extraction, followed by processing through three Conv layers, two LayerNorm layers, a GELU activation function, an SE block, and a DropPath layer, incorporating skip connections to enhance feature recalibration and model robustness.}
    \label{fig07}
\end{figure*}

Compared to the retina with a relatively clear border feature, the conceptual boundary between the macular hole and the macular edema is not clearly defined, and it is difficult to distinguish the macular hole from vitreous opacity and proliferation. Specifically, the upper boundary of the macular hole does not have distinct features, requiring the model to predict the upper boundary by exploiting the information from a wider field of view. Conventional convolutional kernels have a small field of view with most of the focus on local information, and thus can easily result in unclear boundary localization. Therefore, we propose the Harmonic Squeeze-and-Excitation (HSE) Module, which integrates FuGH and Squeeze-and-Excitation (SE) blocks, with a specific structure as shown in Fig.\ref{fig07}.

HSE employs a FuGH module to transform the image analysis from the spatial to the frequency domain, offering an opportunity to suppress irrelevant noise and boost crucial frequency components. As a result, the model working in the frequency domain is more able to differentiate closely related pathological features by filtering out extraneous information and amplifying pertinent details. The operation can be expressed as:
\begin{flalign}
&y_{FuGH}=FuGH(x)&
\end{flalign}
where $x$ and $y_{FuGH}$ are the input feature and output feature of FuGH, respectively.

The SE block is a dynamic recalibration module that enhances the representation ability by adaptively recalibrating channel-wise feature weights \citep{hu2018squeeze}. It models the importance of each feature channel. It guides the network to focus on meaningful characteristics of the current task, by boosting or suppressing different channels in a tailored way. Specifically, by focusing on the feature channels that distinctly represent the macular hole and macular edema, and suppressing feature channels representing proliferation and vitreous opacity, it improves the model's ability to segment small or blurry objects. This process can be expressed as:
\begin{flalign}
&y_{se}=x\odot Sigmoid(FC_2(GELU(FC_1(AvgPool(x)))))&
\end{flalign}
where $x$ is the input feature to the SE block, and $y_{se}$ is the output. $AvgPool$ is an average pooling layer, $FC_1$ and $FC_2$ are two fully connected linear layers, and $GELU$ represents a Gaussian Error Linear Unit. The SE block can be denoted as follows:
\begin{flalign}
&y_{se}=SE(x)&
\end{flalign}

The HSE module integrates a FuGH module and the SE block to constitute an advanced feature extraction and processing unit, as illustrated in Fig.\ref{fig07}. For an input $x$, it goes through the following process:
\begin{flalign}
&x_1 = LN\left(Conv_1 \left( LN(FuGH(x))\right) \right)& \\
&y_{out} = DP\left( SE\left( Conv_3\left( GL\left( LN\left(Conv_2\left(x_1 \right) \right) \right) \right)\right) \right) + x_1&
\end{flalign}
where $y_{out}$ is the output of the HSE module, $Conv_1$, $Conv_2$, and $Conv_3$ correspond to the first, second, and third Conv layers respectively. $x_1$ represents the intermediate quantity of input $x$ before the second Conv layer. $LN$ represents Layer Normalization, a method to normalize the inputs across the features and ensure smooth training. $GL$ stands for Gaussian Error Linear Unit, which is a specific type of activation function employed to introduce non-linearity in the model. $DP$ refers to DropPath, a regularization technique used to prevent overfitting. 

The HSE module, with its dedicated design, captures spatial details and garners frequency domain information. It suppresses interference within the frequency domain while considering inter-channel correlations. This effectively augments the representational power of the network, thereby providing a prospect for enhanced segmentation.

\subsection{Dynamic Weight Composing}
Conventional loss functions are inadequate to accommodate the complexity of segmenting fuzzy boundary objects in 3D sequences. Predominant region-based losses, like the dice loss ($\mathcal{L}_{Dice}$), exhibit suboptimal performance in demarcating boundaries with details, while boundary-focused losses, such as the boundary loss ($\mathcal{L}_{Boundary}$) and the contour loss ($\mathcal{L}_{Contour}$), while focusing on boundary details, struggle with the accurate identification of specific targets like the macular hole or macular edema due to the intricate fundus structure.

To overcome these limitations, we propose an innovative network optimization strategy, namely, the dynamic weight composing (DWC) scheme. Instead of relying on a single optimization objective, DWC dynamically combines several loss functions, each specializing in a distinct segmentation objective, with adaptive weights contingent on the model's training phase. This methodology permits flexible selection and combination of losses and enables crafting a customized overall loss to address the task-specific challenges. By adaptively modulating the weights to align with the optimal loss profile for each training interval, DWC effectively enhances model performance with respect to both the regional and boundary delineation challenges. Based on the concept of DWC, the Dynamic Weight Composing Loss ($\mathcal{L}_{DWC}$) is computed, as an adaptively weighted combination of the focal loss ($\mathcal{L}_{Focal}$), the boundary loss ($\mathcal{L}_{Boundary}$), the dice loss ($\mathcal{L}_{Dice}$), the cross entropy loss ($\mathcal{L}_{CE}$) and the deep ranking loss ($\mathcal{L}_{DeepRanking}$).

\bm{$\mathcal{L}_{Focal}$}:
To concentrate the network's attention on classes with poorer segmentation effects, $\mathcal{L}_{Focal}$ is introduced into the $\mathcal{L}_{DWC}$, as illustrated in the equation that follows:
\begin{flalign}
&input_{ls,c}=\log\frac{exp({input}_c)}{\sum_{i=1}^Cexp({input}_i)}&
\end{flalign}
where $c$ is the index number of categories, $C$ is the total number of categories. 
\begin{flalign}
&\mathcal{L}_{Focal} = \frac{1}{N} \sum_{i=1}^{N} \Bigl[ -\left(1 - e^{input_{ls,c}}\right)^{\gamma} \times {input_{ls,c}} \times target \Bigr]&
\end{flalign}
where $\gamma$ is the gamma value, used to reduce the loss contribution of simple classified samples and enhance the importance of hard-to-classify samples, the $target$ is the result after one-hot encoding of the ground truth, $N$ is the total number of voxels, the batch size multiplied by the number of channels multiplied by the spatial size.

\bm{$\mathcal{L}_{Boundary}$}:
To emphasize the accuracy of segmenting objects with conceptual boundary, $\mathcal{L}_{Boundary}$ is introduced into the $\mathcal{L}_{DWC}$, as explicated in the equation that follows:
\begin{flalign}
&B_{input}=AvgPool3D(I)-I &       \\
&B_{target}=AvgPool3D(T)-T&
\end{flalign}
where $I$ is the input, and $T$ is the target. $B_{input}$ represents the boundary of input, while $B_{target}$ represents the boundary of ground truth.
\begin{flalign}
&\mathcal{L}_{Boundary}=\frac1N\sum_{i=1}^N(B_{input,i}-B_{target,i})^2&
\end{flalign}

\bm{$\mathcal{L}_{Dice}$}:
To ensure the efficacy of the model's overall segmentation and the precision of boundary, $\mathcal{L}_{Dice}$ is incorporated into the $\mathcal{L}_{DWC}$, as delineated in the subsequent equation:
\begin{flalign}
&\mathcal{L}_{\text{Dice}} = \frac{1}{N} \sum_{i=1}^{N} \left( 1 - \frac{2 \times I_i + \epsilon_n}{T_i + P_i + \epsilon_d} \right)&
\end{flalign}
where $p_i$ represents the probability of the prediction belonging to the $i^{th}$ category, $t_i$ is a binary indicator, assigned a value of 1 if the prediction corresponds to the $i^{th}$ category, and 0 otherwise.

\bm{$\mathcal{L}_{CE}$}:
To minimize the discrepancy between the ground truth label distribution and the model's predicted distribution, $\mathcal{L}_{CE}$ is introduced into the $\mathcal{L}_{DWC}$, with the calculation formula as:
\begin{flalign}
&\mathcal{L}_{CE}=-\sum_{i=1}^Ct_i\log(p_i)&
\end{flalign}
where $t_i$ represents the predicted probability that the $i^{th}$ voxel belongs to a certain class, $p_i$ represents the actual label for the $i^{th}$ voxel.

\bm{$\mathcal{L}_{DeepRanking}$}:
To further improve the segmentation performance of the model, $\mathcal{L}_{DeepRanking}$ is introduced into the $\mathcal{L}_{DWC}$. $\mathcal{L}_{DeepRanking}$ can distinguish the distance between positive and negative samples, making positive samples closer to the anchor while negative samples further away from the anchor, with the following formula:
\begin{equation}
\mathcal{L}_{DeepRanking} = \max\left\{0, m+\sum_{i=1}^{N} (p_i-\mu)^2 - \sum_{j=1}^{M} (n_j - \mu)^2\right\}
\end{equation}
where $N$ and $M$ are the numbers of positive and negative samples respectively, and $m$ is a predefined constant known as the margin, it establishes the minimum gap that should exist between the positive and negative samples in the feature space. $p_i$ represents the value of the $i^{th}$ positive sample in the feature space, $n_j$ represents the value of the $j^{th}$ negative sample in the feature space, $\mu$ represents the anchor.

The $\mathcal{L}_{DWC}$ combines the above five loss functions and adjusts the weights of $\mathcal{L}_{Focal}$, $\mathcal{L}_{Boundary}$, $\mathcal{L}_{Dice}$, and $\mathcal{L}_{CE}$ according to different training stages. It also applies $\mathcal{L}_{DeepRanking}$ to the weighted $\mathcal{L}_{Total}$, the formula for weighted calculation is:

\begin{equation}
\begin{aligned}
\mathcal{L}_{Total} = \sum_{i=1}^4 \lambda_{i} \times \mathcal{L}_{i}
\end{aligned}
\end{equation}

Where $i$ ranges from 1 to 4, $\mathcal{L}_i$ corresponds to $\mathcal{L}_{Focal}$, $\mathcal{L}_{Boundary}$, $\mathcal{L}_{Dice}$, and $\mathcal{L}_{CE}$ respectively, while  $\lambda_i$ represents their respective weighting coefficients.

The effectiveness of the DWC has been validated in the 3D segmentation task of fundus OCT. It can be foreseen that DWC can also provide help to researchers in various tasks in other fields.

\section{Experiments and results}
\subsection{Datasets}
The DEFN is evaluated on the OIMHS \citep{ye2023oimhs} dataset. A subset of OIMHS serves as the test set, while the remainder, along with data enhanced by SDi from CARS-30k, forms the training and validation sets.


\subsubsection{Training dataset}
The OCT Image Macular Hole Segmentation (OIMHS) dataset is a publicly accessible retinal OCT segmentation repository, distinguished by its composition of 3,859 B-scan SD-OCT images featuring macular holes. This dataset is unparalleled in its volume among the publicly available datasets in its domain. It includes OCT scans from 119 patients diagnosed with macular holes, resulting in a total of 125 sequences, each encompassing 19 to 73 slices. The images manifest in varied resolutions: 220 images at 384×496 pixels, 3,002 images at 512×496 pixels, and 637 images at 768×496 pixels. The regions are categorized into four classes: macular hole, retina, macular edema, and choroid, with the choroid category not considered in this experiment.

\subsubsection{Comprehensive archive for retinal segmentation}
A multicentre dataset from the retinal OCT scans of patients was collected, named Comprehensive Archive for Retinal Segmentation (CARS-30k), constitutes a pivotal augmentation to the existing image datasets in the realm of retinal research. The CARS-30k dataset encompasses 30,684 B-scan SD-OCT images, carefully annotated, depicting instances of macular edema, sourced from Spectralis HRA (Heidelberg Engineering, Germany). 
The scans are carefully selected to exclude scans with severe artifacts or significantly diminished signal intensity, which could obscure the retinal interface. The resulting collection contains 1,476 independent scan sequences from 137 patients diagnosed with DME, each sequence comprising between 13 to 49 slices, with a resolution of 496x512 pixels each.

A rigorous annotation process was followed, involving five radiologists with specialized expertise in retinal imaging, followed by a validation step executed by two senior retinal specialists to ascertain the accuracy and consistency of the annotations. 

Some statistics of the two datasets are given in Table \ref{table01}.

\newcolumntype{L}{>{\fontsize{4}{5}\selectfont}l}

\begin{table}[!h]
\centering
\caption{Detailed information of OIMHS and CARS-30k, including patient number, sequence number, and slice number.}
\scalebox{2}{
\renewcommand{\arraystretch}{0.5}
\begin{tabular}{L|L|L}
\toprule
Dataset name     & OIMHS & CARS-30k                     \\
\midrule
Patients number  & 119       & 137                          \\
Sequences number & 125       & 1476                         \\
Slices number    & 3859      & 30684                        \\
\bottomrule
\end{tabular}
}
\label{table01}
\end{table}

\subsection{Evaluation metrics}
To comprehensively evaluate the proposed approach, several metrics are employed, including Mean Intersection over Union ($MIoU$), Dice coefficient ($Dice$), Average Symmetric Surface Distance ($ASSD$), Hausdorff Distance ($HD$), the 95th percentile Hausdorff Distance ($HD95$), and the Adjusted Rand Index ($AdjRand$). These metrics offer a multifaceted assessment of segmentation accuracy, similarity, and consistency, facilitating a robust analysis of performance.

$MIoU$, one of the main metrics for the evaluation of image segmentation, quantitatively assesses the overlap between predicted and true regions, offering insight into the precision of semantic segmentation. It is mathematically represented as:
\begin{equation}
MIoU= \frac{1}{N} \sum_{i=1}^{N} \frac{|V_{pred,i} \cap V_{true,i}|}{|V_{pred,i} \cup V_{true,i}|} \times100\%
\end{equation}
where $N$ is the total number of categories, $V_{pred,i}$ is the region of the $i^{th}$ predicted category, and $V_{true,i}$ is the region of the $i^{th}$ true category.

$Dice$ evaluates the similarity between two samples, serving as a crucial metric in image segmentation to quantify the congruence between predicted and true regions:
\begin{equation}
Dice = \frac1N\sum_{i=1}^N\frac{2 \times |V_{pred,i} \cap V_{true,i}|}{|V_{pred,i}| + | V_{true,i}|}\times100\%
\end{equation}

$ASSD$ gauges the average distance between the true and predicted boundaries, thereby providing a measure of the segmentation's spatial accuracy:
\begin{equation}
\begin{aligned}
ASSD &= \frac{1}{N} \sum_{i=1}^{N} \Bigg( \frac{1}{|S_{true,i}|} \sum_{s \in S_{true,i}}\mathrm{d_{min}}(s, S_{pred,i}) \\
&\qquad + \frac{1}{|S_{pred,i}|} \sum_{s \in S_{pred,i}} \mathrm{d_{min}}(s, S_{true,i}) \Bigg)
\end{aligned}
\end{equation}
where $S_{\mathrm{true},i}$ and $S_{\mathrm{pred},i}$ are the true and predicted voxel sets of the $i^{th}$ surface, respectively. $\mathrm{d_{min}}(s, S)$ represents the minimum distance from voxel $s$ to the surface $S$. $|S_{\mathrm{true}, i}|$ and $|S_{\mathrm{pred}, i}|$ are the total number of voxels in the true and predicted voxel sets of the $i^{th}$ surface, respectively.

$HD$ quantifies the maximum distance between two point sets, in this context, between the predicted and true boundaries, thus reflecting the extremities of segmentation error:
\begin{equation}
HD = \max\left\{\max_{a \in S_{true}} \min_{b \in S_{pred}} \mathrm{d}(a, b), \max_{b \in S_{pred}} \min_{a \in S_{true}} \mathrm{d}(b, a)\right\}
\end{equation}
where $\mathrm{d}(a, b)$ is the distance between voxel $a$ in $S_{true}$ and voxel $b$ in $S_{pred}$.

$HD95$, a variant of $HD$, employs the 95th percentile of all point pair distances to mitigate the impact of outliers, offering a more robust measure against extreme values.

AdjRand, ranging from [-1, 1], measures the similarity between two data partitions, with 1 indicating perfect agreement, 0 denoting no better than chance, and -1 representing complete disagreement, thus evaluating the segmentation consistency:
\begin{equation}
\begin{aligned}
AdjRand = \frac{
    \sum_{ij} \binom{n_{ij}}{2} - \left[ \sum_i \binom{a_i}{2} \times \sum_j \binom{b_j}{2} \right] / \binom{n}{2}
}{
    \frac{1}{2} \left[ \sum_i \binom{a_i}{2} + \sum_j \binom{b_j}{2} \right] - \left[ \sum_i \binom{a_i}{2} \times \sum_j \binom{b_j}{2} \right] / \binom{n}{2}
}
\end{aligned}
\end{equation}
where $n_{ij}$ is the number of data points that are shared between the $i^{th}$ predicted cluster and the $j^{th}$ true cluster. $a_i$ is the number of data points in the predicted cluster $i$. $b_j$ is the number of data points in the true cluster $j$. $n$ is the total number of data points. $\binom{n}{2}$ calculates the number of combinations of two elements chosen from $n$ elements.

\subsection{Experimental setting}
All experiments were executed in a uniform hardware and software environment. The hardware comprised a desktop computer equipped with eight NVIDIA 4090 graphics cards, an Intel E5-2690V4 CPU, and 256GB RAM. The software foundation was Python 3.9 and PyTorch 2.0.0.

For model training, the AdamW optimizer was utilized with a learning rate of 0.0001. Model weights were initialized randomly, with all models undergoing an identical number of epochs during the pre-training and fine-tuning phases. The input size for the OIMHS and CARS-30k datasets was 96$\times$96$\times$96 voxels with a batch size of 8. Data augmentation was performed leveraging the SDi strategy, enhancing training data through flipping, rotating, translating, scaling, adding Gaussian noise, and applying random histogram transformations.

\subsection{Stochastic defect injection pre-training}
In this section, we examine the performance of the proposed method on the OIMHS dataset following training on the CARS-30k dataset with two SDi data augmentation techniques. The outcomes are compared with those from existing methods.

\begin{table}[!t]
  \centering
    \caption{Segmentation results employing the isolated injection method, comparing the proposed DEFN, DEFN+$\mathcal{L}_{DWC}$ and prior classic models. The evaluation spans four classes: All (Average across all categories), MH (Macular Hole), ME (Macular Edema), and RA (Retina). The best values for each metric are highlighted in red, the second-best in blue, and the values of our model are bolded.}
\resizebox{\linewidth}{9cm}{
\small
\renewcommand{\arraystretch}{0.5}
\begin{tabular}{lccccccc}
\toprule
 \multicolumn{1}{l}{} & \multicolumn{6}{c}{All}                           \\
 
\cmidrule(lr){2-7}
\makebox[0.25\textwidth][l]{} & \makebox[0.14\textwidth][c]{\textit{MIoU}} & \makebox[0.14\textwidth][c]{\textit{Dice}}  & \makebox[0.14\textwidth][c]{\textit{ASSD}}
     & \makebox[0.14\textwidth][c]{\textit{HD}} & \makebox[0.14\textwidth][c]{\textit{HD95}} & \makebox[0.14\textwidth][c]{\textit{AdjRand}}             \\
\midrule
3D UX-NET \citep{lee20223d}                    & 60.40±14.28                                 & 68.75±15.97                                 & 15.58±62.93                                  & {\color[HTML]{0000FF} 118.17±301.68}           & 42.08±157.17                                 & 68.11±15.87                                 \\
nnFormer   \citep{zhou2023nnformer}            & 54.91±12.91                                 & 62.85±15.33                                 & 19.18±72.30                                  & 129.05±211.07                                  & 40.56±135.20                                 & 62.13±15.18                                 \\
3D U-Net   \citep{cciccek20163d}               & 57.53±13.81                                 & 65.51±15.91                                 & 14.63±33.30                                  & 136.01±348.79                                  & 41.15±181.76                                 & 64.83±15.82                                 \\
SegResNet  \citep{myronenko20193d}             & 60.81±13.76                                 & 70.07±15.50                                 & 9.79±20.20                                   & 181.43±320.77                                  & 37.65±162.15                                 & 69.40±15.45                                 \\
\bottomrule
\end{tabular}
}

  \label{table02}
\end{table}

\begin{table*}[!t]
  \centering
  \caption{Segmentation results employing the comprehensive injection method, comparing the proposed DEFN, DEFN+$\mathcal{L}_{DWC}$ and prior classic models. The evaluation spans four classes: All (Average across all categories), MH (Macular Hole), ME (Macular Edema), and RA (Retina). The best values for each metric are highlighted in red, the second-best in blue, and the values of our model are bolded.}
\resizebox{\linewidth}{9cm}{
\small
\renewcommand{\arraystretch}{0.5}
\begin{tabular}{lccccccc}
\toprule
 \multicolumn{1}{l}{} & \multicolumn{6}{c}{All}                           \\
 
\cmidrule(lr){2-7}
\makebox[0.25\textwidth][l]{} & \makebox[0.14\textwidth][c]{\textit{MIoU}} & \makebox[0.14\textwidth][c]{\textit{Dice}}  & \makebox[0.14\textwidth][c]{\textit{ASSD}}
     & \makebox[0.14\textwidth][c]{\textit{HD}} & \makebox[0.14\textwidth][c]{\textit{HD95}} & \makebox[0.14\textwidth][c]{\textit{AdjRand}}             \\

\midrule
3D UX-NET \citep{lee20223d}                & 66.28±13.83                                 & 77.05±15.44                                 & 5.01±4.80                                  & 133.14±252.72                                 & 31.54±156.40                                 & 76.36±15.32                                 \\
nnFormer    \citep{zhou2023nnformer}           & 61.23±12.90                                 & 72.49±14.87                                 & 4.95±5.28                                  & 128.16±268.41                                 & 30.30±207.76                                 & 71.74±14.74                                 \\
3D U-Net    \citep{cciccek20163d}              & 65.19±14.38                                 & 76.00±15.84                                 & 6.05±6.10                                  & 145.94±371.01                                 & 32.76±195.83                                 & 75.26±15.79                                 \\
SegResNet   \citep{myronenko20193d}            & 65.10±13.76                                 & 76.17±15.53                                 & 5.43±5.53                                  & 154.90±417.70                                 & 33.37±189.19                                 & 75.42±15.42                                 \\
\bottomrule
\end{tabular}
}
  \label{table03}
\end{table*}

\subsubsection{Isolated injection training}
\begin{figure*}[!h]
    \centering
    \includegraphics[width=0.8\linewidth]{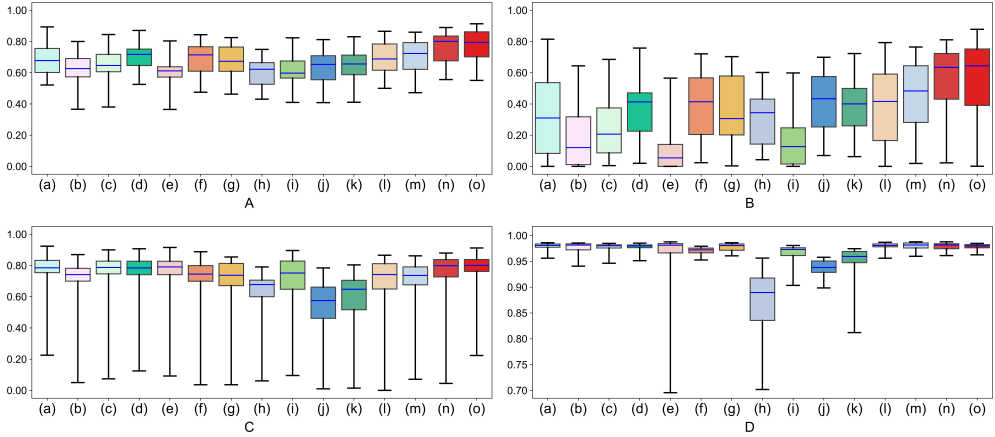}
    \caption{Box plot showcasing the Dice index from training results using the isolated injection method.  The categories are represented as follows: table A (Average across all classes), table B (Macular Hole), table C (Macular Edema), and table D (Retina).  On the x-axis, the models are labeled as follows: (a) 3D UX-Net;  (b) nnFormer;  (c) 3D U-Net;  (d) SegResNet;  (e) Swin UNETR;  (f) TransBTS;  (g) UNETR;  (h) DeepResUNet;  (i) ResUNet;  (j) HighRes3DNet;  (k) MultiResUNet;  (l) SegCaps;  (m) V-Net: (n) DEFN;  (o) DEFN+$\mathcal{L}_{DWC}$.}
    \label{fig08}
\end{figure*}

The isolated injection strategy is first employed for training on the CARS-30k dataset and subsequent testing on the OIMHS dataset. The results are shown in Table.\ref{table02} and Fig.\ref{fig08}. When employing the usual DiceCE Loss ($\mathcal{L}_{DiceCE}$), DEFN leads to satisfactory results across all categories. This is evidenced by the competitive scores in the $MIoU$, $Dice$, and $AdjRand$ metrics. The macular hole category, in particular, witnesses a notable enhancement of over 10\% in both $MIoU$ and $Dice$ metrics compared to previous methodologies, substantiating the efficacy of the FuGH in terms of deep information extraction and denoising, and in feature extraction by S3DSA. On top of that, employing $\mathcal{L}_{DWC}$ along with DEFN results in superior performance across all metrics, with the biggest advance in the macular category, where $MIoU$ and $Dice$ scores are increased by 2.9\% and 2.59\%, respectively. This improvement is attributable to the synergistic action of $\mathcal{L}_{Focal}$ and $\mathcal{L}_{Boundary}$ integrated within $\mathcal{L}_{DWC}$, focusing more on classes with suboptimal segmentation outcomes and ambiguous boundaries, thus refining the segmentation performance of easily confounded small targets and boundaries. As a result, the proposed methodology outranks in all metrics for both macular hole and macular edema categories. For all metrics, the mean scores across all categories outperform other methods with large margins, underscoring the network's proficiency in both comprehensive and detailed segmentation levels. A few segmentation samples are presented in Fig.\ref{fig10}, to visually illustrate the difference in output by SoTA methods, and showcase the precise delineation of the retinal structures by the proposed methods.

\begin{figure*}[!h]
    \centering
    \includegraphics[width=0.8\linewidth]{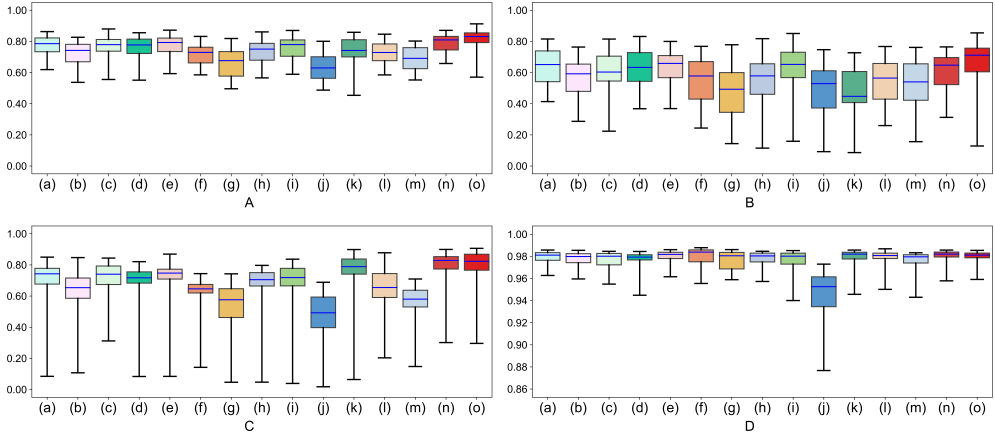}
    \caption{Box plot showcasing the Dice index from training results using the comprehensive injection method.  The categories are represented as follows: table A (Average across all classes), table B (Macular Hole), table C (Macular Edema), and table D (Retina).  On the x-axis, the models are labeled as follows: (a) 3D UX-Net;  (b) nnFormer;  (c) 3D U-Net;  (d) SegResNet;  (e) Swin UNETR;  (f) TransBTS;  (g) UNETR;  (h) DeepResUNet;  (i) ResUNet;  (j) HighRes3DNet;  (k) MultiResUNet;  (l) SegCaps;  (m) V-Net: (n) DEFN;  (o) DEFN+$\mathcal{L}_{DWC}$.}
    \label{fig09}
\end{figure*}

\subsubsection{Comprehensive injection training}
\begin{figure*}[p]
    \centering
    \includegraphics[width=\linewidth]{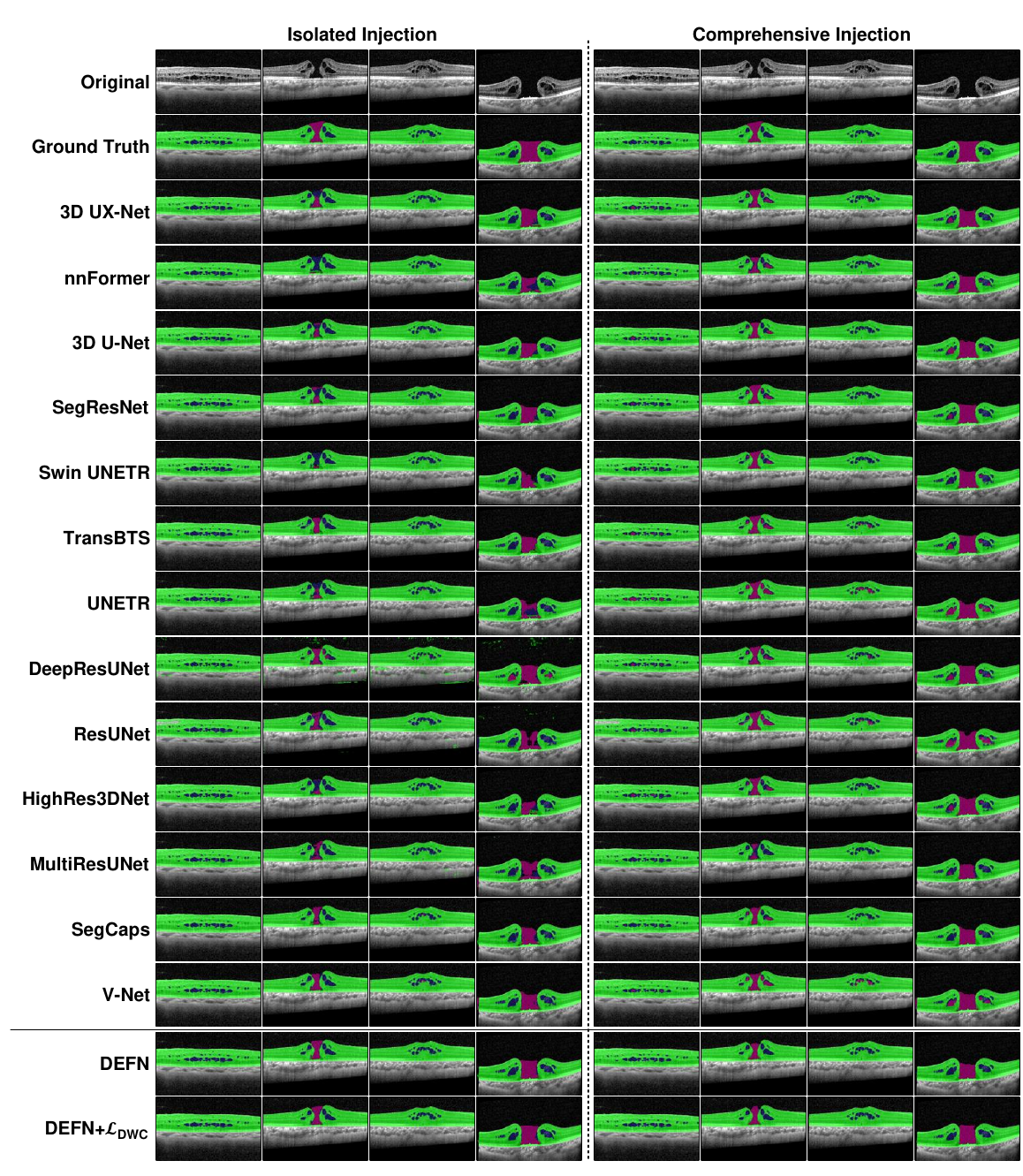}
    \caption{A few segmentation results using the SDi method, with a comparison between the proposed DEFN, DEFN+$\mathcal{L}_{DWC}$, and the existing state-of-the-art models, selected from the OIMHS dataset. The left four images showcase the results of the isolated injection strategy, while the right four depict the outcomes of the comprehensive injection strategy. The segmentation result of the retina is shown in green, macular edema in blue, and macular hole in red.}
    \label{fig10}
\end{figure*}

The comprehensive injection strategy is also applied for data preprocessing and model training on the CARS-30k dataset, followed by the evaluation on OIMHS dataset. The results are presented in Table \ref{table03} and Fig. \ref{fig09}. Without $\mathcal{L}_{DWC}$, the DEFN already exhibits superior performance in $MIoU$ and $Dice$ for macular edema and retina categories. This is particularly notable in the macular edema category, where relative to the most proficient competing models,$MIoU$ and $Dice$ are increased by 3.92\% and 3.54\%, respectively. The enhancement is primarily attributed to the integration of the FuGH module alongside the HSE module, which effectively enhances DEFN's capability to comprehend data regularities and suppress interference noise and unnecessary feature channels. The incorporation of the $\mathcal{L}_{DWC}$ optimization strategy leads to further improvements across all metrics, with the mean $Dice$ scores across all categories achieving a pinnacle of 80.92\%. Moreover, the scores in $ASSD$, $HD$, and $HD95$ all exhibit noticeable advancements. This improvement is largely due to the $\mathcal{L}_{DWC}$ optimization strategy, which guides the model to first focus on enhancing regional accuracy, and subsequently shifting it towards the precision of segmentation boundaries. A few selected samples of segmentation results are depicted in Fig.\ref{fig10}.

\subsection{Stochastic defect injection fine-tuning on OIMHS dataset}
In this section, we examine the performance of the proposed method on the OIMHS dataset following initial pre-training on the CARS-30k dataset employing two distinct SDi data augmentation strategies. This process entails further training on the OIMHS dataset. The corresponding evaluation results are contrasted with those of the SoTA methodologies.

\subsubsection{Isolated injection fine-tuning}
\begin{figure*}[!h]
    \centering
    \includegraphics[width=0.8\linewidth]{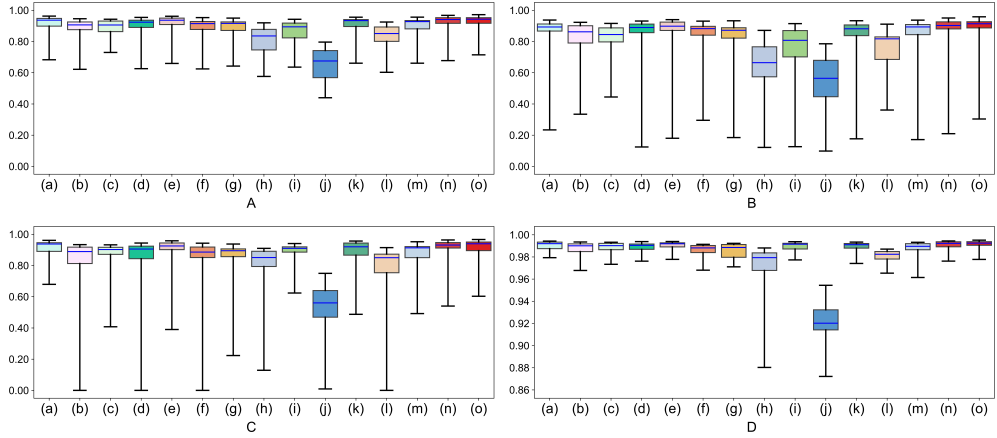}
    \caption{Box plot showcasing the Dice index from fine-tuning results after isolated injection training. The categories are represented as follows: table A (Average across all classes), table B (Macular Hole), table C (Macular Edema), and table D (Retina).  On the x-axis, the models are labeled as follows: (a) 3D UX-Net;  (b) nnFormer;  (c) 3D U-Net;  (d) SegResNet;  (e) Swin UNETR;  (f) TransBTS;  (g) UNETR;  (h) DeepResUNet;  (i) ResUNet;  (j) HighRes3DNet;  (k) MultiResUNet;  (l) SegCaps;  (m) V-Net: (n) DEFN;  (o) DEFN+$\mathcal{L}_{DWC}$.}
    \label{fig11}
\end{figure*}

Utilizing the isolated injection strategy for data augmentation and pre-training on the CARS-30k dataset, followed by fine-tuning on the OIMHS dataset, the DEFN showed better results in all aspects, as shown in Table.\ref{table04} and Fig.\ref{fig11}. Without employing the $\mathcal{L}_{DWC}$ optimization strategy, DEFN alone generates good segmentation accuracy for macular holes, retina, and the average scores across all categories, with superior $MIoU$, $Dice$, and $AdjRand$ scores. Specifically, in the most challenging category of macular holes, DEFN scores 78.52\% and 86.81\% in $MIoU$ and $Dice$, respectively, with a margin of 0.93\%  and 0.62\% compared to the SoTA results. The incorporation of the $\mathcal{L}_{DWC}$ optimization strategy leads to further enhancements across all metrics. For instance, in the macular hole category, $MIoU$ and $Dice$ are increased by an additional 1.39\% and 1.11\%, respectively, $ASSD$, $HD$, and $HD95$ are improved as well. This validates the benefit of incorporating $\mathcal{L}_{DWC}$ in optimizing both overall segmentation accuracy and detailed boundary accuracy. As a result, DEFN attained impressive scores across nearly all evaluation criteria for macular holes, macular edema, and retinal categories, showcasing its competitive edge. Selected segmentation results are depicted in Fig.\ref{fig13}.

\begin{table*}[!h]
  \centering
  \caption{Segmentation results of fine-tuning after isolated injection training, comparing the proposed DEFN, DEFN+$\mathcal{L}_{DWC}$ and prior classic models. The evaluation spans four classes:  All (Average across all categories), MH (Macular Hole), ME (Macular Edema), and RA (Retina). The best values for each metric are highlighted in red, the second-best in blue, and the values of our model are bolded.}
\resizebox{\linewidth}{9cm}{
\small
\renewcommand{\arraystretch}{0.5}
\begin{tabular}{lccccccc}
\toprule
 \multicolumn{1}{l}{} & \multicolumn{6}{c}{All}                           \\
 
\cmidrule(lr){2-7}
\makebox[0.25\textwidth][l]{} & \makebox[0.14\textwidth][c]{\textit{MIoU}} & \makebox[0.14\textwidth][c]{\textit{Dice}}  & \makebox[0.14\textwidth][c]{\textit{ASSD}}
     & \makebox[0.14\textwidth][c]{\textit{HD}} & \makebox[0.14\textwidth][c]{\textit{HD95}} & \makebox[0.14\textwidth][c]{\textit{AdjRand}}             \\
\midrule
3D UX-NET ~{} \citep{lee20223d}                    & 85.44±17.18                                 & 91.34±17.58                                 & 2.24±5.41                                  & 103.31±120.88                                & 12.03±112.02                                 & 91.01±17.55                                 \\
nnFormer   ~{} \citep{zhou2023nnformer}            & 80.55±17.22                                 & 87.72±17.92                                 & 1.74±4.12                                  & 102.52±396.50                                & 10.01±86.35                                  & 87.30±17.90                                 \\
3D U-Net   ~{} \citep{cciccek20163d}               & 81.26±16.62                                 & 88.54±17.05                                 & 3.70±4.65                                  & 155.41±533.76                                & 26.46±197.83                                 & 88.15±17.04                                 \\
SegResNet  ~{} \citep{myronenko20193d}             & 83.31±17.34                                 & 89.52±18.18                                 & 2.69±11.79                                 & 162.22±361.54                                & 7.52±80.96                                   & 89.16±18.12                                 \\
\bottomrule
\end{tabular}
}
  \label{table04}
\end{table*}

\begin{table*}[!h]
  \centering
  \caption{Segmentation results of fine-tuning after comprehensive injection training, comparing the proposed DEFN, DEFN+$\mathcal{L}_{DWC}$ and prior classic models. The evaluation spans four classes:  All (Average across all categories), MH (Macular Hole), ME (Macular Edema), and RA (Retina). The best values for each metric are highlighted in red, the second-best in blue, and the values of our model are bolded.}

\resizebox{\linewidth}{9cm}{
\small
\renewcommand{\arraystretch}{0.5}
\begin{tabular}{lccccccc}
\toprule
 \multicolumn{1}{l}{} & \multicolumn{6}{c}{All}                           \\
 
\cmidrule(lr){2-7}
\makebox[0.25\textwidth][l]{} & \makebox[0.14\textwidth][c]{\textit{MIoU}} & \makebox[0.14\textwidth][c]{\textit{Dice}}  & \makebox[0.14\textwidth][c]{\textit{ASSD}}
     & \makebox[0.14\textwidth][c]{\textit{HD}} & \makebox[0.14\textwidth][c]{\textit{HD95}} & \makebox[0.14\textwidth][c]{\textit{AdjRand}}             \\
\midrule
3D UX-NET ~{} \citep{lee20223d}                    & 83.91±16.03                                 & 90.61±16.83                                 & 1.56±2.24                                  & 110.23±308.81                                & 5.89±39.85                                 & 90.25±16.77                                 \\
nnFormer   ~{} \citep{zhou2023nnformer}            & 70.75±17.75                                 & 79.61±18.96                                 & 4.00±12.64                                 & 49.58±307.11                                 & 16.21±129.44                               & 79.01±18.98                                 \\
3D U-Net   ~{} \citep{cciccek20163d}               & 82.76±16.81                                 & 89.56±17.24                                 & 1.62±4.98                                  & 103.89±330.29                                & 6.84±71.83                                 & 89.17±17.22                                 \\
SegResNet  ~{} \citep{myronenko20193d}             & 82.96±17.23                                 & 89.36±17.90                                 & 2.58±8.08                                  & 108.66±594.89                                & 10.26±103.68                               & 88.96±17.85                                 \\
\bottomrule
\end{tabular}
}
  \label{table05}
\end{table*}

\subsubsection{Comprehensive injection fine-tuning}
\begin{figure*}[!h]
    \centering
    \includegraphics[width=0.8\linewidth]{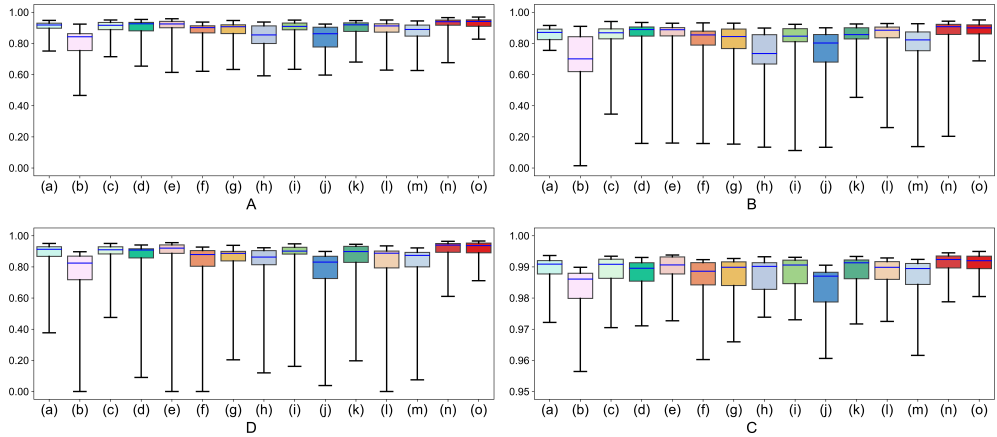}
    \caption{Box plot showcasing the Dice index from fine-tuning results after comprehensive injection training. The categories are represented as follows: table A (Average across all classes), table B (Macular Hole), table C (Macular Edema), and table D (Retina).  On the x-axis, the models are labeled as follows: (a) 3D UX-Net;  (b) nnFormer;  (c) 3D U-Net;  (d) SegResNet;  (e) Swin UNETR;  (f) TransBTS;  (g) UNETR;  (h) DeepResUNet;  (i) ResUNet;  (j) HighRes3DNet;  (k) MultiResUNet;  (l) SegCaps;  (m) V-Net: (n) DEFN;  (o) DEFN+$\mathcal{L}_{DWC}$.}
    \label{fig12}
\end{figure*}

\begin{figure*}[p]
    \centering
    \includegraphics[width=1\linewidth]{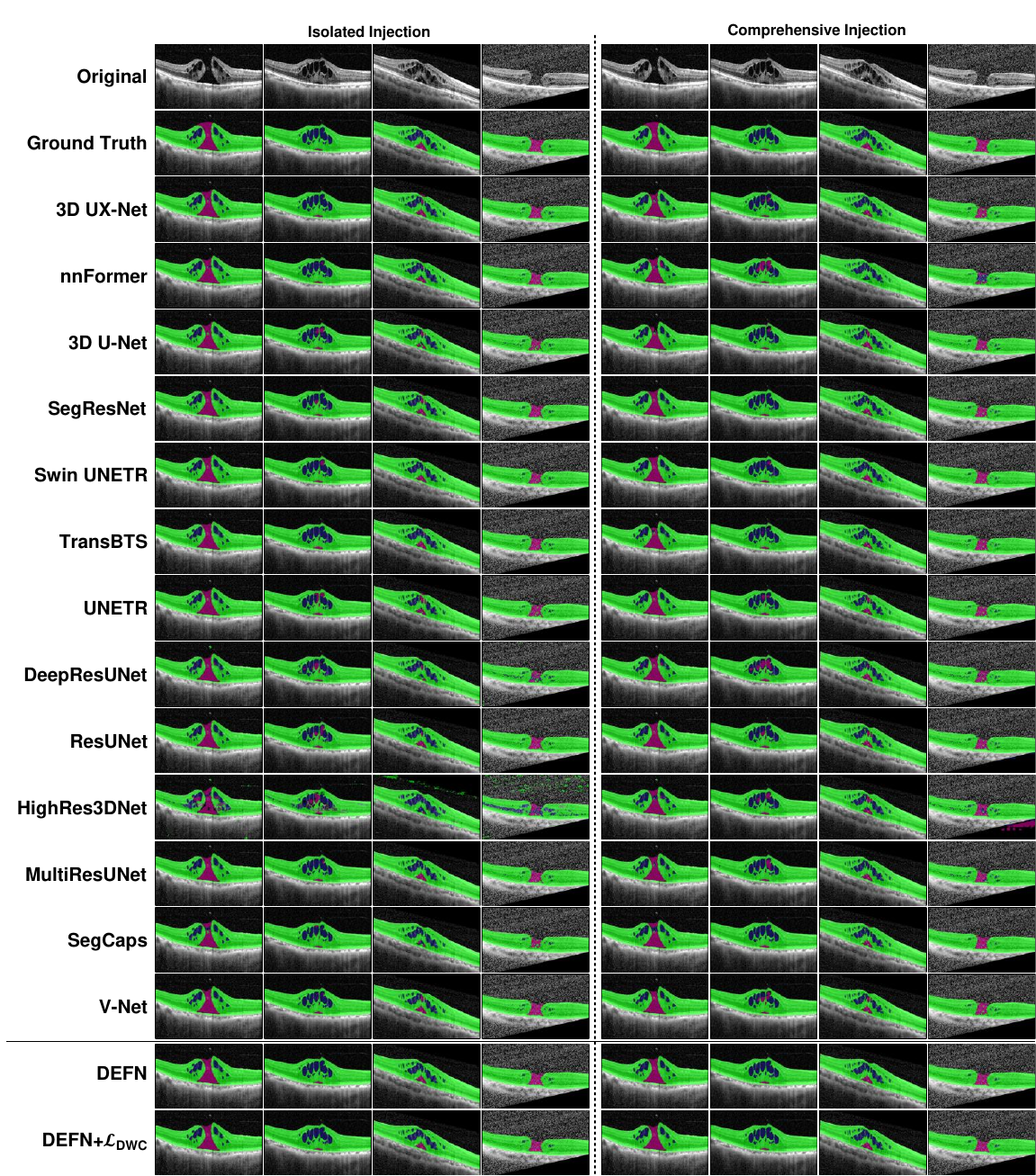}
    \caption{A few segmentation results of fine-tuning after SDi training, with a comparison between the proposed DEFN, DEFN+$\mathcal{L}_{DWC}$, and previous classic models on the OIMHS dataset. The left four images showcase the results of the isolated injection strategy, while the right four depict the outcomes of the comprehensive injection strategy. The segmentation result of the retina is shown in green, the segmentation result of the macular edema is shown in blue and the segmentation result of the macular hole is shown in red.}
    \label{fig13}
\end{figure*}

This experiment utilizes the comprehensive injection strategy for data augmentation and pre-training on the CARS-30k dataset, followed by fine-tuning on the OIMHS dataset. DEFN again yields satisfactory results, as presented in Table \ref{table06} and Fig. \ref{fig12}. A comparative analysis with existing segmentation models demonstrates the DEFN's superior performance across all metrics. The DEFN, with a common loss, manifested a superior segmentation capability across all categories, as evidenced by its scores in $MIoU$, $Dice$, $ASSD$, and $AdjRand$. In this experiment, DEFN exhibits proficiency in discerning between macular holes and macular edema, showcasing its robustness against substantial noise and defects. This robustness is attributed to the FuGH module's exemplary noise mitigation capacity and the S3DSA module's spatial information extraction. Furthermore, the strategic selection of feature channels in the HSE module enables DEFN to predict the upper boundary of the macular hole category with a $Dice$ score of 86.77\%, surpassing the most adept alternative models by 1.19\%. Clear advantages are also shown in the ASSD, HD, and HD95 metrics. The integration of the $\mathcal{L}_{DWC}$ optimization strategy leads to a further improvement in all metrics, with a remarkable 1.54\% enhancement in the $Dice$ score for the macular hole category and a clear reduction in the $ASSD$, $HD$, $HD95$ distances. Selected samples of segmentation outcomes are graphically represented in Fig. \ref{fig13}.

\begin{table*}[!t]
    \centering
    \caption{Ablation study on the backbone and the proposed methods, including DEFN (DEFN backbone), HSE (Harmonic Squeeze-and-Excitation module), FuGH (Fourier Group Harmonics module), IMHI (isolated macular hole injection), CMHI (comprehensive injection) and DWC (Dynamic Weight Composing Loss). The best values for each metric are highlighted in red, while the second-best values are highlighted in blue.}
\resizebox{\linewidth}{7cm}{
\tiny
\renewcommand{\arraystretch}{0.5}
\begin{tabular}{cccccccccccc}
\toprule
\multicolumn{1}{l}{} & \multicolumn{1}{l}{} & \multicolumn{1}{l}{} & \multicolumn{1}{l}{} & \multicolumn{1}{l}{} & \multicolumn{1}{l}{} & \multicolumn{6}{c}{All}                                                 \\
\cmidrule(lr){1-6}  \cmidrule(lr){7-12}
DEFN                 & HSE                  & FuGH                 & IMHI                 & CMHI                 & DWC                  & \textit{MIoU}                               & \textit{Dice}                               & \textit{ASSD}                              & \textit{HD}                                                      & \textit{HD95}                                & \textit{AdjRand}                            \\
\midrule
\checkmark                    &                      &                      &                      &                      &                      & 84.24±17.39                        & 90.20±18.12                        & 1.49±6.91                         & 88.16±527.05                                            & 6.60±77.35                          & 89.84±18.07                        \\
\checkmark                    & \checkmark                    &                      &                      &                      &                      & 85.62±17.63                        & 91.20±18.08                        & 1.42±6.93                         & 88.96±505.39                                            & 6.39±76.46                          & 90.87±18.04                        \\
\checkmark                    & \checkmark                    & \checkmark                    &                      &                      &                      & 85.83±17.38                        & 91.43±17.88                        & 1.70±7.49                         & 83.17±341.72                                            & 8.88±100.12                         & 91.11±17.84                        \\
\checkmark                    & \checkmark                    & \checkmark                    & \checkmark                    &                      &                      & 86.56±17.34                        & 92.05±17.76                        & 1.15±6.86                         & 29.85±99.24                                             & 5.06±69.50                          & 91.75±17.71                        \\
\checkmark                    & \checkmark                    & \checkmark                    &                      & \checkmark                    &                      & 86.44±17.49                        & 91.91±17.81                        & 1.35±6.32                         & 93.55±562.18                                            & 6.37±77.02                          & 91.60±17.78                        \\
\checkmark                    & \checkmark                    & \checkmark                    &                      &                      & \checkmark                    & 86.60±17.28                        & 92.08±17.82                        & 1.29±7.38                         & 43.32±313.34                                            & 5.04±69.89                          & 91.79±17.78                        \\
\checkmark                    & \checkmark                    & \checkmark                    & \checkmark                    &                      & \checkmark                    & {\color[HTML]{FF0000} 87.41±17.23} & {\color[HTML]{FF0000} 92.66±17.58} & {\color[HTML]{0000FF} 1.03±5.08}  & {\color[HTML]{0000FF} 28.59±97.19}                      & {\color[HTML]{0000FF} 4.98±72.83}   & {\color[HTML]{FF0000} 92.37±17.54} \\
\checkmark                    & \checkmark                    & \checkmark                    &                      & \checkmark                    & \checkmark                    & {\color[HTML]{0000FF} 86.95±16.60} & {\color[HTML]{0000FF} 92.61±17.01} & {\color[HTML]{FF0000} 0.40±0.77}  & {\color[HTML]{FF0000} 28.38±70.05}                      & {\color[HTML]{FF0000} 2.54±10.37}   & {\color[HTML]{0000FF} 92.30±16.97} \\
\bottomrule
\toprule
                     &                      &                      &                      &                      &                      & \multicolumn{6}{c}{MH}                                                                                                                                                                                                                           \\
\cmidrule(lr){1-6}  \cmidrule(lr){7-12}
DEFN                 & HSE                  & FuGH                 & IMHI                 & CMHI                 & DWC                  & \textit{MIoU}                               & \textit{Dice}                               & \textit{ASSD}                              & \textit{HD}                                                      & \textit{HD95}                                & \textit{AdjRand}                            \\
\midrule
\checkmark                    &                      &                      &                      &                      &                      & 76.07±22.49                        & 85.20±23.08                        & 3.58±45.79                        & 159.27±1346.95                                          & 16.49±673.78                        & 85.07±23.09                        \\
\checkmark                    & \checkmark                    &                      &                      &                      &                      & 77.39±22.28                        & 86.15±22.68                        & 3.19±45.40                        & 127.54±1897.66                                          & 15.97±672.56                        & 86.02±22.69                        \\
\checkmark                    & \checkmark                    & \checkmark                    &                      &                      &                      & 77.96±22.96                        & 86.39±23.46                        & 3.80±48.75                        & 113.57±1883.15                                          & 18.40±793.55                        & 86.27±23.47                        \\
\checkmark                    & \checkmark                    & \checkmark                    & \checkmark                    &                      &                      & 78.41±22.57                        & 86.77±22.90                        & 2.84±45.59                        & 27.43±705.31                                            & 11.11±604.81                        & 86.66±22.91                        \\
\checkmark                    & \checkmark                    & \checkmark                    &                      & \checkmark                    &                      & 78.52±22.76                        & 86.81±22.91                        & 3.29±40.96                        & 149.51±1677.10                                          & 15.76±672.56                        & 86.68±22.93                        \\
\checkmark                    & \checkmark                    & \checkmark                    &                      &                      & \checkmark                    & 78.18±23.00                        & 86.52±23.78                        & 3.19±50.25                        & 50.19±1272.39                                           & 11.06±607.37                        & 86.40±23.79                        \\
\checkmark                    & \checkmark                    & \checkmark                    & \checkmark                    &                      & \checkmark                    & {\color[HTML]{FF0000} 79.91±21.83} & {\color[HTML]{0000FF} 87.92±21.51} & {\color[HTML]{0000FF} 2.48±30.60} & {\color[HTML]{0000FF} 26.01±727.94}                     & {\color[HTML]{0000FF} 10.93±637.41} & {\color[HTML]{0000FF} 87.81±21.51} \\
\checkmark                    & \checkmark                    & \checkmark                    &                      & \checkmark                    & \checkmark                    & {\color[HTML]{0000FF} 79.47±16.82} & {\color[HTML]{FF0000} 88.31±16.93} & {\color[HTML]{FF0000} 0.79±2.31}  & {\color[HTML]{FF0000} 21.10±216.65}                     & {\color[HTML]{FF0000} 4.90±74.43}   & {\color[HTML]{FF0000} 88.19±16.92} \\
\bottomrule
\toprule
                     &                      &                      &                      &                      &                      & \multicolumn{6}{c}{ME}                                                                                                                                                                                                                           \\
\cmidrule(lr){1-6}  \cmidrule(lr){7-12}
DEFN                 & HSE                  & FuGH                 & IMHI                 & CMHI                 & DWC                  & \textit{MIoU}                               & \textit{Dice}                               & \textit{ASSD}                              & \textit{HD}                                                      & \textit{HD95}                                & \textit{AdjRand}                            \\
\midrule
\checkmark                    &                      &                      &                      &                      &                      & 78.90±25.06                        & 86.53±26.23                        & 0.73±3.35                         & 67.18±639.12                                            & 2.28±7.54                           & 86.36±26.20                        \\
\checkmark                    & \checkmark                    &                      &                      &                      &                      & 81.57±24.30                        & 88.50±24.27                        & 0.88±4.21                         & 103.80±780.24                                           & {\color[HTML]{0000FF} 2.16±7.70}    & 88.36±24.23                        \\
\checkmark                    & \checkmark                    & \checkmark                    &                      &                      &                      & 81.61±22.04                        & 88.96±21.09                        & 1.17±7.19                         & 115.34±658.36                                           & 7.21±266.93                         & 88.81±21.06                        \\
\checkmark                    & \checkmark                    & \checkmark                    & \checkmark                    &                      &                      & 83.18±19.31                        & 90.34±18.31                        & 0.47±1.64                         & 46.85±446.58                                            & 3.03±23.68                          & 90.20±18.29                        \\
\checkmark                    & \checkmark                    & \checkmark                    &                      & \checkmark                    &                      & 82.69±20.42                        & 89.88±19.11                        & 0.63±1.75                         & {\color[HTML]{FF0000} 87.53±704.40}                     & 2.30±8.18                           & 89.73±19.10                        \\
\checkmark                    & \checkmark                    & \checkmark                    &                      &                      & \checkmark                    & {\color[HTML]{0000FF} 83.49±18.20} & {\color[HTML]{0000FF} 90.67±17.66} & 0.55±1.52                         & 55.67±494.78                                            & 3.02±23.43                          & {\color[HTML]{0000FF} 90.53±17.63} \\
\checkmark                    & \checkmark                    & \checkmark                    & \checkmark                    &                      & \checkmark                    & {\color[HTML]{FF0000} 84.14±19.11} & {\color[HTML]{FF0000} 90.98±18.28} & {\color[HTML]{0000FF} 0.47±1.51}  & {\color[HTML]{0000FF} 48.72±418.89} & 2.96±22.46                          & {\color[HTML]{FF0000} 90.84±18.26} \\
\checkmark                    & \checkmark                    & \checkmark                    &                      & \checkmark                    & \checkmark                    & \multicolumn{1}{l}{83.28±18.48}    & 90.47±17.69                        & {\color[HTML]{FF0000} 0.29±0.30}  & 48.76±492.19                                            & {\color[HTML]{FF0000} 1.67±1.60}    & 90.33±17.68                        \\

\bottomrule
\toprule

\multicolumn{1}{l}{} & \multicolumn{1}{l}{} & \multicolumn{1}{l}{} & \multicolumn{1}{l}{} & \multicolumn{1}{l}{} & \multicolumn{1}{l}{} & \multicolumn{6}{c}{RA}                                                                                                                                                             \\
\cmidrule(lr){1-6}  \cmidrule(lr){7-12}
DEFN                 & HSE                  & FuGH                 & IMHI                 & CMHI                 & DWC                  & \textit{MIoU}                               & \textit{Dice}                               & \textit{ASSD}                              & \textit{HD}                                                      & \textit{HD95}                                & \textit{AdjRand}                            \\
\midrule
\checkmark                    &                      &                      &                      &                      &                      & 97.74±17.56                        & 98.85±17.75                        & 0.16±0.10                         & 38.03±1009.48                                           & 1.04±0.31                           & 98.07±17.62                        \\
\checkmark                    & \checkmark                    &                      &                      &                      &                      & 97.91±17.59                        & 98.94±17.76                        & 0.17±0.20                         & 35.54±529.72                                            & 1.04±0.31                           & 98.23±17.64                        \\
\checkmark                    & \checkmark                    & \checkmark                    &                      &                      &                      & 97.92±17.59                        & 98.95±17.76                        & 0.14±0.09                         & 20.59±119.56                                            & 1.04±0.31                           & 98.24±17.65                        \\
\checkmark                    & \checkmark                    & \checkmark                    & \checkmark                    &                      &                      & 98.10±17.62                        & 99.04±17.78                        & 0.14±0.10                         & 15.27±108.29                                            & 1.04±0.31                           & 98.39±17.67                        \\
\checkmark                    & \checkmark                    & \checkmark                    &                      & \checkmark                    &                      & 98.10±17.62                        & 99.04±17.78                        & {\color[HTML]{0000FF} 0.13±0.09}  & {\color[HTML]{0000FF} 43.61±1029.30}                    & 1.04±0.31                           & 98.39±17.67                        \\
\checkmark                    & \checkmark                    & \checkmark                    &                      &                      & \checkmark                    & {\color[HTML]{0000FF} 98.15±17.63} & {\color[HTML]{0000FF} 99.06±17.78} & {\color[HTML]{FF0000} 0.13±0.09}  & 24.10±662.69                                            & 1.04±0.31                           & {\color[HTML]{0000FF} 98.43±17.68} \\
\checkmark                    & \checkmark                    & \checkmark                    & \checkmark                    &                      & \checkmark                    & {\color[HTML]{FF0000} 98.20±17.64} & {\color[HTML]{FF0000} 99.09±17.79} & 0.13±0.10                         & {\color[HTML]{FF0000} 11.03±9.24}                       & {\color[HTML]{FF0000} 1.04±0.31}    & {\color[HTML]{FF0000} 98.47±17.69} \\
\checkmark                    & \checkmark                    & \checkmark                    &                      & \checkmark                    & \checkmark                    & 98.09±17.62                        & 99.03±17.78                        & 0.14±0.10                         & 15.28±107.36                                            & {\color[HTML]{0000FF} 1.04±0.31}    & 98.38±17.67                       \\
\bottomrule
\end{tabular}
}
    \label{table06}
\end{table*}

\subsection{Ablation study}
\begin{figure*}[!h]
    \centering
    \includegraphics[width=0.8\linewidth]{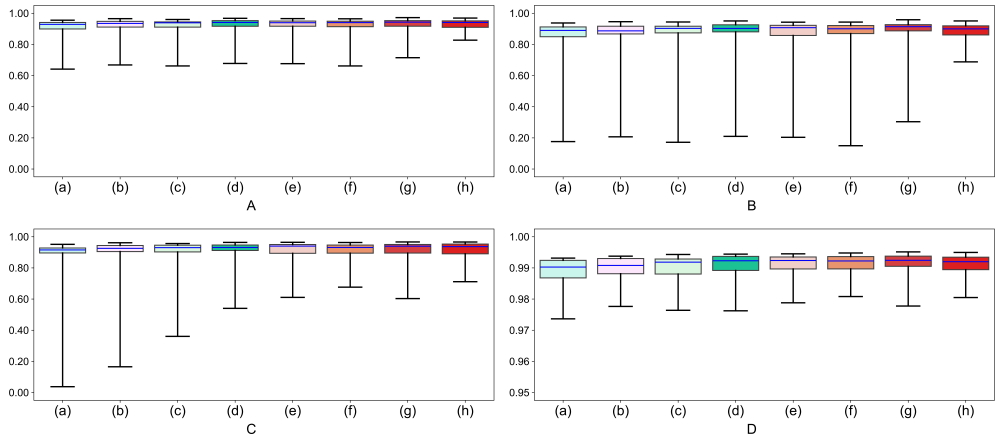}
    \caption{Box plot showcasing the Dice index from ablation study. The categories are denoted as follows: table A (Average across all classes), table B (Macular Hole), table C (Macular Edema), and table D (Retina). On the x-axis, the models are labeled as follows: (a) DEFN; (b) DEFN+HSE; (c) DEFN+HSE+FuGH; (d) DEFN+HSE+FuGH+IMHI; (e) DEFN+HSE+FuGH+CMHI; (f) DEFN+HSE+FuGH+DWC; (g) DEFN+HSE+FuGH+IMHI+DWC; (h) DEFN+HSE+FuGH+CMHI+DWC.}
    \label{fig14}
\end{figure*}

To corroborate the efficacy and reliability of the proposed approach, we executed ablation studies, focusing on the influence of the HSE branch, FuGH module, two SDi strategies, and $\mathcal{L}_{DWC}$ optimization strategy on model efficacy. 
Results delineated in Table.\ref{table06} and Fig.\ref{fig14}, unequivocally establish the indispensability of the modules of DEFN. The omission of the HSE branch precipitates an apparent degradation across all performance indicators, with the average $MIoU$ and $Dice$ scores across all categories diminishing by 1.38\% and 1\%, respectively. This substantiates the pivotal role of the HSE branches in fortifying segmentation capabilities. Concurrently, both the FuGH module and two SDi expansion strategies, benefit model segmentation performance. The incorporation of the FuGH module facilitates a discernible increment of 0.57\% in the $MIoU$ score within the macular hole category, and the performance of the two SDi expansion strategies ultimately converges, exhibiting a negligible 0.04\% divergence in the $Dice$ metric for the macular hole category without employing the $\mathcal{L}_{DWC}$ optimization strategy. This observation underscores a consistent performance enhancement across differing SDi strategies, suggesting that the choice of injection strategy, while contributory, does not critically dictate the efficacy of model performance enhancement. Moreover, the deployment of the $\mathcal{L}_{DWC}$ optimization strategy, independent of the SDi expansion strategy, notably enhances segmentation accuracy, as evidenced by a 0.65\% augmentation in the average $Dice$ score across all categories and a decrement in the average $ASSD$ from 1.7 to 1.29. These outcomes collectively affirm the efficacy of the $\mathcal{L}_{DWC}$ strategy in ameliorating regional segmentation and edge detail delineation.

\subsection{3D reconstruction and quantitative index}
To elevate the value of clinical applications, an accompanying fundus 3D reconstruction scheme to the proposed segmentation method is introduced. This scheme enables accurate, real-time 3D modeling of the fundus, providing clinicians with an intuitive tool for patient condition assessment. Built upon the Early Treatment Diabetic Retinopathy Study (ETDRS) grid, it delivers comprehensive and precise quantitative metrics. These metrics empower clinicians to make more accurate and reasoned evaluations of patient conditions, thus streamlining the disease monitoring process. Moreover, during the treatment phase, these quantitative indices offer reliable guidance to clinicians, boosting decision-making efficiency and treatment effectiveness.

\begin{figure*}[!t]
    \centering
    \includegraphics[width=0.9\linewidth]{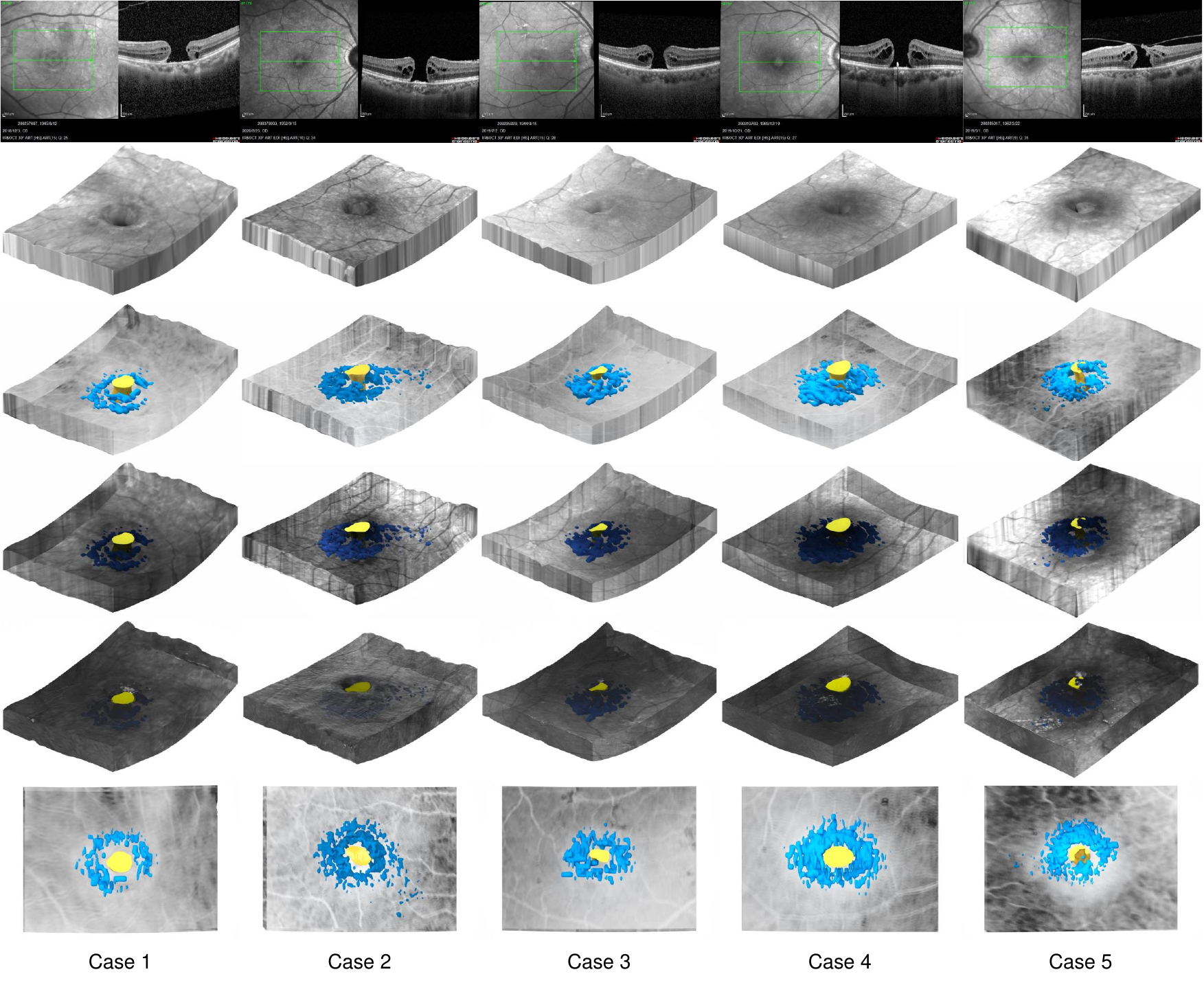}
    \caption{Five cases are presented with their original images and reconstruction outcomes. The first row exhibits the original images for each case. Rows two to five show the reconstructions based on four different rendering styles, while the sixth row provides a top view of the reconstruction results. Within the reconstructions, yellow regions indicate macular holes and blue regions signify macular edema.}
    \label{fig15}
\end{figure*}

\subsubsection{3D reconstruction}
The proposed 3D reconstruction method facilitates the automatic generation of 3D fundus modeling subsequent to segmentation, incorporating four distinct rendering styles to accommodate various observational perspectives and emphasize specific areas of interest as necessitated by clinical requirements. Fig. \ref{fig15} illustrates the original images alongside outcomes derived from the four rendering styles, in addition to a top-view representation of the reconstructed models. The reconstruction process involves three principal stages: initially, three-dimensional voxels are generated based on segmentation outcomes; subsequent voxel smoothing enhances model fidelity; and finally, comprehensive 3D models are produced. Performance evaluation of the method reveals that a typical sequence's 3D reconstruction on a single NVIDIA GeForce RTX 4090 GPU requires merely 33 seconds, and the usage time is stable for input sequences of any length.

\subsubsection{Quantitative index calculation}
To furnish clinicians with more precise and clinically relevant quantitative metrics, the ETDRS grid, extensively utilized within the field of ophthalmology, was employed. Fig. \ref{fig16}(a) delineates the grid, comprising nine distinct regions and three concentric circles:

\begin{figure}[!ht]
    \centering
    \includegraphics[width=0.4\linewidth]{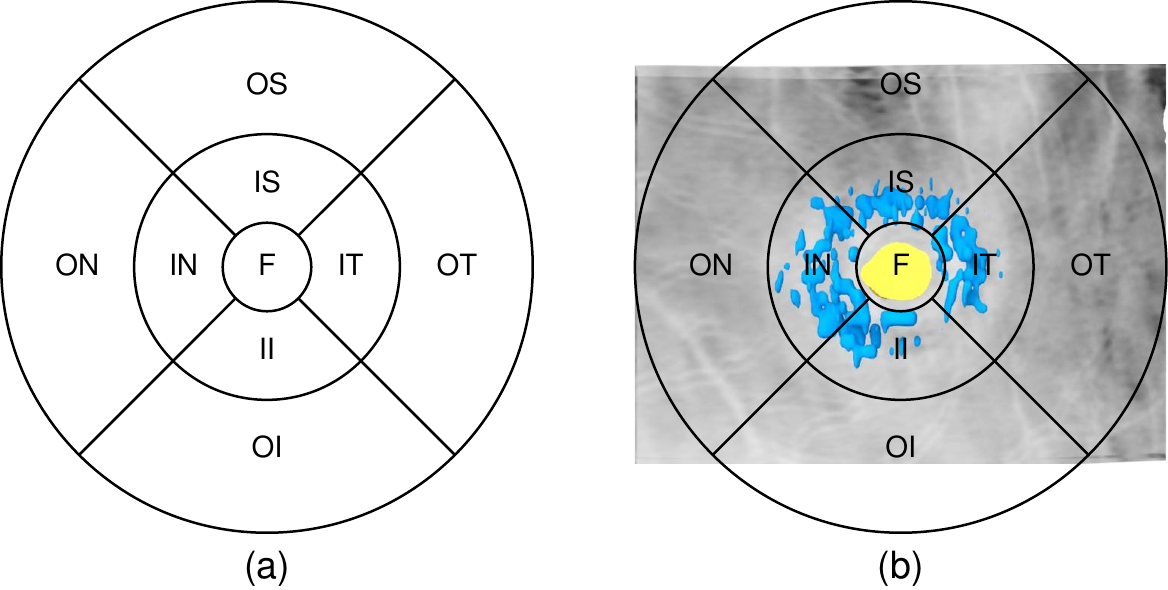}
    \caption{The ETDRS grid is shown as (a), which contains 9 sectors: F (Foveal), IS (Inner Superior), II (Inner Inferior), IT (Inner Temporal), IN (Inner Nasal), OS (Outer Superior), OI (Outer Inferior), OT (Outer Temporal), ON (Outer Nasal); and three concentric circles: central circle, intermediate circle, outer circle. (b) represents the schematic diagram of the ETDRS superimposed on the reconstructed retina after proportional scaling.}
    \label{fig16}
\end{figure}

\textbf{F (Foveal)}: represents the central depressed part of the macula, which is the source of human central vision.

\textbf{IS (Inner Superior)}: represents the region above the macula center.

\textbf{II (Inner Inferior)}: represents the region below the macula center.

\textbf{IT (Inner Temporal)}: represents the region toward the nasal side of the macula center.

\textbf{IN (Inner Nasal)}: represents the region toward the temporal side of the macula center.

\textbf{OS (Outer Superior)}: represents the region outside the upper region of the macula center.

\textbf{OI (Outer Inferior)}: represents the region outside the lower region of the macula center.

\textbf{OT (Outer Temporal)}: represents the region outside the nasal side region of the macula center.

\textbf{ON (Outer Nasal)}: represents the region outside the temporal side region of the macula center.

\textbf{Central circle}: Centered on the central depression (macula center), with a diameter of 1 \textbf{optic disc} (The optic disc represents a modestly elevated, circular or elliptical region located on the retina of the eyeball, playing an indispensable role in the diagnosis and management of a spectrum of ophthalmic disorders. Clinicians frequently utilize the diameter of the optic disc as a benchmark. In most cases, 1 optic disc is equivalent to 1 millimeter. Within the ophthalmic practice, this metric often serves as a convenient unit to describe the extent of pathological changes on the retina.). This area is also referred to as the "central zone" or "macular zone".

\textbf{Intermediate circle}: Centered on the central depression, with a diameter of 3 optic discs. This circle surrounds the central circle, with a width of 1 optic disc. This area is also referred to as the "inner circle" or "intermediate zone".

\textbf{Outer circle}: Centered on the central depression, with a diameter of 6 optic discs. This circle surrounds the intermediate circle, with a width of 3 optic discs. This area is also called the "outer circle" or "peripheral zone".

\begin{table}[!t]
\centering
\caption{Diagram depicting volume calculations for different sections according to the ETDRS zoning chart, and the volumes inside the intermediate circle and the outer circle, where IC represents the intermediate circle (diameter of 3mm), OC represents the outer circle (diameter of 6mm), MH represents the macular hole, ME represents the macular edema, RA represents the retina. These five cases correspond to the examples in Fig.\ref{fig15}.}

\newcolumntype{C}{>{\fontsize{20}{20}\selectfont}c} 
\setlength{\tabcolsep}{2pt} 
\small
\resizebox{\textwidth}{!}{ 
\renewcommand{\arraystretch}{1.5} 
\begin{tabular}{c|ccccccccccc|ccccccccccc|ccccccccccc}
\toprule
 & \multicolumn{11}{c|}{MH} & \multicolumn{11}{c|}{ME} & \multicolumn{11}{c}{RA} \\
\cmidrule(lr){2-34}
 & IC & OC & F & IN & IT & IS & II & ON & OS & OT & OI & IC & OC & F & IN & IT & IS & II & ON & OS & OT & OI & IC & OC & F & IN & IT & IS & II & ON & OS & OT & OI \\
\midrule
Case 1 & 0.056 & 0.056 & 0.056 & 0.000 & 0.000 & 0.000 & 0.000 & 0.000 & 0.000 & 0.000 & 0.000 & 0.066 & 0.066 & 0.003 & 0.015 & 0.021 & 0.015 & 0.012 & 0.000 & 0.000 & 0.000 & 0.000 & 2.213 & 7.132 & 0.172 & 0.492 & 0.536 & 0.505 & 0.509 & 1.450 & 0.958 & 1.556 & 0.955 \\
Case 2 & 0.035 & 0.035 & 0.035 & 0.000 & 0.000 & 0.000 & 0.000 & 0.000 & 0.000 & 0.000 & 0.000 & 0.155 & 0.158 & 0.022 & 0.045 & 0.030 & 0.011 & 0.047 & 0.000 & 0.001 & 0.003 & 0.000 & 2.521 & 7.596 & 0.254 & 0.557 & 0.582 & 0.560 & 0.568 & 1.488 & 0.988 & 1.667 & 0.931 \\
Case 3 & 0.020 & 0.020 & 0.020 & 0.000 & 0.000 & 0.000 & 0.000 & 0.000 & 0.000 & 0.000 & 0.000 & 0.088 & 0.088 & 0.017 & 0.031 & 0.020 & 0.006 & 0.016 & 0.000 & 0.000 & 0.000 & 0.000 & 2.393 & 7.213 & 0.257 & 0.535 & 0.544 & 0.526 & 0.532 & 1.433 & 0.936 & 1.520 & 0.930 \\
Case 4 & 0.067 & 0.067 & 0.067 & 0.000 & 0.000 & 0.000 & 0.000 & 0.000 & 0.000 & 0.000 & 0.000 & 0.246 & 0.247 & 0.013 & 0.099 & 0.051 & 0.020 & 0.062 & 0.001 & 0.000 & 0.000 & 0.000 & 2.690 & 7.776 & 0.254 & 0.632 & 0.612 & 0.569 & 0.623 & 1.529 & 0.969 & 1.596 & 0.993 \\
Case 5 & 0.033 & 0.033 & 0.033 & 0.000 & 0.000 & 0.000 & 0.000 & 0.000 & 0.000 & 0.000 & 0.000 & 0.143 & 0.144 & 0.012 & 0.023 & 0.042 & 0.004 & 0.062 & 0.000 & 0.000 & 0.001 & 0.000 & 2.454 & 7.386 & 0.248 & 0.516 & 0.588 & 0.533 & 0.570 & 1.410 & 0.906 & 1.691 & 0.920 \\
\bottomrule
\end{tabular}
}
\label{table07}
\end{table}

Clinicians may leverage the delineated regions within the ETDRS grid to articulate the precise location and severity of lesions with greater accuracy, facilitating the prognosis of potential visual complications and the formulation of tailored treatment regimens. For instance, lesions situated at the macula center (F region) might necessitate urgent interventions to avert further visual deterioration. While lesions within the peripheral zone might not demand immediate action they should undergo consistent monitoring to inhibit propagation to vital regions.

Consequently, the quantitative metrics derived from the ETDRS grid offer substantial clinical utility. Upon the completion of fundus 3D reconstruction, the ETDRS grid was applied to the constructed model for regional categorization, enabling the quantification of the retina, macular hole, and macular edema volumes within each specified region and circle. This quantitative analysis permits clinicians to assess the macular hole's severity through data-driven insights across diverse regions, serving as a pivotal reference in the selection of treatment protocols and bearing significant clinical relevance.

Specifically, the operation process is as follows: First, as shown in Fig.\ref{fig16}(b), scale the ETDRS grid to match the reconstructed ocular structure, ensuring that they are adjusted in size according to the actual physical space ratio. Then superimpose them by taking the macula as the center. On this basis, we separately calculate the volume of the retina, the macular hole, and the macular edema in each ETDRS region and within each concentric circle. Table.\ref{table07} summarizes the calculation results of the five cases in Fig.\ref{fig15}.

\section{Clinical impact}
In this study, our task was to develop a method for the 3D reconstruction and quantification of indistinct-boundary objects. We developed a framework for high-accuracy segmentation, reconstruction and quantification of fundus structures, and can be used clinically. The foundational segmentation model DEFN exhibits decent efficacy in segmenting critical fundus components, including macular holes and macular edema. Leveraging this model, we achieved high-precision 3D reconstruction of ocular structures. Our approach diverges from the majority of preceding investigations by not solely advancing algorithmic developments but also by addressing the pragmatic requirements of clinical applications. For instance, whereas previous research predominantly focused on the accuracy of image segmentation, our study extends to consider the real-world implications for surgical decision-making and disease prognosis, enhancing its clinical applicability. The core contribution of this study is to provide highly precise three-dimensional modeling and quantitative indicators for doctors in the surgical decision-making process, which can optimizes surgical programs and allows data-based prediction and evaluation of treatment effects in the postoperative stage. For the macular hole, common treatment methods include vitrectomy surgery, which usually involves the injection of silicone oil or sulfur hexafluoride (C3F8) filler during surgery. Using the effective segmentation capabilities of the DEFN model and high-precision 3D reconstruction in this study, clinicians can accurately ascertain the volume and precise localization of the macular hole before vitrectomy. This is vital for determining the appropriate type and quantity of filler, given the direct correlation between the macular hole's volume and the requisite filler volume. During the surgical intervention, high-definition 3D imagery offers clinicians a vivid, three-dimensional perspective, thereby enhancing the precision of surgical executions. Quantitative metrics derived postoperatively for the macular hole are instrumental in monitoring the progression of disease recovery and assessing treatment efficacy, such as through evaluating the stability of silicone oil or C3F8. This enables a more precise understanding of the postoperative recovery landscape and the potential necessity for pharmacological intervention for both clinicians and patients.

In summary, this study proposed DEFN for accurate delineation of indistinct-boundary or clear small object, and utilizes segmentation results for 3D reconstruction and the quantification of various indicators in alignment with clinical practices. This provides support for a range of clinical activities, encompassing diagnosis, surgical intervention, and postoperative monitoring. Through appropriate fine-tuning, DEFN has the potential to extend its application to other complex diseases such as various forms of macular degeneration or retinal detachment, providing clinicians with a more comprehensive suite of diagnostic and treatment tools.

\section{Conclusions}
In this paper, we propose the Dual-Encoder Fourier Group Harmonics Network (DEFN) for indistinct-boundary medical objects, which provides a framework for addressing some of the most persistent challenges in image segmentation—sparse data, noise interference, and indistinct boundaries. It extends to a rigorous exposition of the stochastic defect injection (SDi) and dynamic weight composing (DWC), which are integral to refining the model's operability under constrained data environments and varied noise conditions. The DEFN architecture, underpinned by the Fourier Group Harmonics (FuGH), operates within the frequency domain, facilitating a meticulous dissection of the noise components that traditionally obfuscate the segmentation process. By transmuting the spatial data into the frequency domain using FFT, the model is better equipped to segregate and suppress extraneous noise frequencies, thereby amplifying the signal integrity of pertinent features. Such capabilities are crucial for the meticulous three-dimensional reconstruction required in preoperative evaluations and intraoperative navigations. Additionally, proposed Simplified 3D Spatial Attention (S3DSA) module prioritizes critical spatial features within complex images to enhance sensitivity to subtle anatomical variations in regions of interest. The Harmonic Squeeze-and-Excitation (HSE) module further refines the model's channel-wise feature recalibration capabilities by optimizing the processing of frequency-domain information. Further, the introduction of the SDi technique addresses the lack of representative samples in medical datasets, especially for rare or subtle conditions. The DWC mechanism introduces a novel paradigm in model training by adjusting the influence of different loss functions throughout training. This adaptability allows the DEFN to finely tune its parameters to the specificities of the dataset and its inherent challenges. The DWC systematically modulates the emphasis on various aspects of the loss landscape, ensuring that the model's learning trajectory is optimally aligned with the evolving dynamics of the training process. Experiments conducted on the public dataset OIMHS demonstrate our method achieves SOTA performance. Both the Dice and HD metrics indicate good outcomes, reflecting the competitive capabilities of our method in overall segmentation and boundary detection. 
Future study aims to dissect and evaluate the performance impacts of diverse architectural configurations post-Fourier transformation, particularly focusing on how these configurations interact with the transformed frequency domain representations. A central objective will be to methodically analyze the efficacy of different convolutional strategies that operate in the frequency domain, comparing group convolutions, depth-wise separable convolutions, and dilated convolutions, among others. It is anticipated that a more granular understanding of frequency domain dynamics can be achieved, potentially leading to significant improvements in the model's ability to isolate and enhance medical-relevant signals amidst noise.

\section*{Acknowledgements}
This work was supported by the National Natural Science Foundation of China (No. 62071415), Natural Science Foundation of Zhejiang Province (No. LY21H160031), Medical Health Science and Technology Project of Zhejiang Provincial Health Commission Grants (No.2024KY111 and No. 2024KY1105).



\bibliographystyle{elsarticle-num}
\bibliography{refs}






\end{document}